\documentclass[useAMS,usenatbib]{mn2e}

\usepackage{ulem}
\usepackage[dvips]{graphicx}
\usepackage{graphics}
\usepackage{lscape}
\usepackage{amssymb}
\usepackage{mathrsfs}
\input{epsf}

\newcommand\Lsun{{\,L_\odot}}
\newcommand\Msun{{\,M_\odot}}

\title[AGN host galaxy colour gradients]
      {Host galaxy colour gradients and accretion disc obscuration in AEGIS $z \sim 1$ X-ray-selected active galactic nuclei}

\author[C.\ M.\ Pierce et al.]{
  \parbox[t]{\textwidth}{
  C.\ M.\ Pierce,$^{1,2}$\thanks{E-mail: christina.pierce@physics.gatech.edu}
  J.\ M.\ Lotz,$^{3}$\thanks{Leo Goldberg Fellow}
  S.\ Salim,$^{3}$
  E.\ S.\ Laird,$^{4}$
  A.\ L.\ Coil,$^{5}$
  K.\ Bundy,$^{6}$ \\
  C.\ N.\ A.\ Willmer,$^{7}$
  D.\ J.\ V.\ Rosario,$^{8}$
  J.\ R.\ Primack$^{9}$ and
  S.\ M.\ Faber$^{8}$
  }
\vspace*{5pt} \\
$^{1}$Department of Physics, University of California, Santa Cruz, 1156 High Street, Santa Cruz, CA 95064, USA \\
$^{2}$School of Physics, Georgia Institute of Technology, 837 State Street, Atlanta, GA 30332-0430, USA \\
$^{3}$National Optical Astronomical Observatories, 950 N.\ Cherry Avenue, Tucson, AZ 85719, USA \\
$^{4}$Astrophysics Group, Imperial College London, Blackett Laboratory, Prince Consort Rd., London SW7 2AW, UK \\
$^{5}$Department of Physics, University of California, San Diego, CA 92093, USA \\
$^{6}$Department of Astronomy, University of California, Berkeley, CA 94720, USA \\
$^{7}$Steward Observatory, University of Arizona, 933 North Cherry Avenue, Tucson, AZ 85721, USA \\
$^{8}$UCO/Lick Observatory; Department of Astronomy and Astrophysics, University of California, Santa Cruz, 1156 High Street, \\ \hspace{0.9mm} Santa Cruz, CA 95064, USA \\
$^{9}$Santa Cruz Institute of Particle Physics, University of California, Santa Cruz, 1156 High Street, Santa Cruz, CA 95064, USA \\
}

\begin{document}

\date{Accepted 2010 June 3. Received 2010 June 2; in original form 2010 January 20}

\pagerange{\pageref{firstpage}--\pageref{lastpage}} \pubyear{2010}

\maketitle

\label{firstpage}

%%%%%%%%%%%%%%%%%%%%%%%%%%%%%
% Abstract
%%%%%%%%%%%%%%%%%%%%%%%%%%%%%

\begin{abstract}
We describe the effect of AGN light on host galaxy optical and UV-optical colours, as determined from X-ray-selected AGN host galaxies at $z \sim 1$, and compare the AGN host galaxy colours to those of a control sample matched to the AGN sample in both redshift and stellar mass. We identify as X-ray-selected AGNs $8.7^{+4}_{-3}$ per cent of the red-sequence control galaxies, $9.8\pm{3}$ per cent of the blue-cloud control galaxies, and $14.7^{+4}_{-3}$ per cent of the green-valley control galaxies. The nuclear colours of AGN hosts are generally {\it bluer} than their outer colours, while the control galaxies exhibit {\it redder} nuclei. AGNs in blue-cloud host galaxies experience less X-ray obscuration, while AGNs in red-sequence hosts have more, which is the reverse of what is expected from general considerations of the interstellar medium. Outer and integrated colours of AGN hosts generally agree with the control galaxies, regardless of X-ray obscuration, but the nuclear colours of unobscured AGNs are typically much bluer, especially for X-ray luminous objects. Visible point sources are seen in many of these, indicating that the nuclear colours have been contaminated by AGN light and that obscuration of the X-ray radiation and visible light are therefore highly correlated. Red AGN hosts are typically slightly bluer than red-sequence control galaxies, which suggests that their stellar populations are slightly younger. We compare these colour data to current models of AGN formation. The unexpected trend of less X-ray obscuration in blue-cloud galaxies and more in red-sequence galaxies is problematic for all AGN feedback models, in which gas and dust is thought to be removed as star formation shuts down. A second class of models involving radiative instabilities in hot gas is more promising for red-sequence AGNs but predicts a larger number of point sources in red-sequence AGNs than is observed. Regardless, it appears that multiple AGN models are necessary to explain the varied AGN host properties discussed in the current work. Finally, we find that integrated optical and UV-optical colours are not strongly affected by X-ray-selected AGNs except in rare cases ($<10$ per cent) where the AGN is very luminous, unobscured, and/or visible as a point source.
\end{abstract}


%%%%%%%%%%%%%%%%%%%%%%%%%%%%%
% Introduction
%%%%%%%%%%%%%%%%%%%%%%%%%%%%%
\section{Introduction}\label{intro}
Galaxy colours can tell us much about the stellar populations and star formation history (SFH) of a galaxy. Of particular interest are the possible connections between the SFH and possible energetic feedback related to black hole growth. However, in order to properly interpret the measured colours of a galaxy we must understand the origins of the colours. Radiation associated with an accretion disc around active galactic nuclei (AGNs) produces large amounts of blue light, causing potentially unexpected effects on the measured colours of the host galaxies.

Star formation, AGN feedback and possible connections between them have been discussed by several authors in recent years (e.g.\ Silk \& Rees 1998; Granato et al.\ 2004; Hopkins et al.\ 2005a, 2005b, 2006, 2008a, 2008b; Scannapieco, Silk, \& Bouwens 2005; Bower et al.\ 2006; Cattaneo et al.\ 2006; Croton et al.\ 2006; Dekel \& Birnboim 2006; Ciotti \& Ostriker 2007). We will compare the results presented here to a few specific scenarios. One such scenario described AGNs caused by interactions or mergers between gas-rich disc galaxies having similar masses (Hopkins et al.\ 2008a, 2008b). Another explored black hole growth initiated by instabilities in isolated giant elliptical galaxies (Ciotti \& Ostriker 2007). Both of these scenarios suggest timelines for various observable features, such as nuclear obscuration and the shutting down of star formation. We also compare our results to a scenario that described how feedback from low-luminosity AGNs may prevent additional star formation (Croton et al.\ 2006) following an initial burst associated with, for example, a merger or interaction.

Ciotti \& Ostriker (2007) described simulations related to the circumstances surrounding significant growth of supermassive black holes (SMBHs) found in giant elliptical galaxies. In their simulations, gas emitted from central stars led to radiative instabilities and a collapse of metal-rich gas in the nuclear regions of the galaxy. New star formation claimed about half of the gas, and about half was ejected from the nucleus; less than 1 per cent of the gas contributed to the growth of the central SMBH. Ciotti \& Ostriker (2007) found that both the AGN and the starburst were heavily obscured (Compton-thick; $N_{H} > 10^{24}$ cm$^{-2}$) during this stage and probably only observable in far-infrared bands. Radiative feedback from the AGN then caused the expansion of a central hot bubble, first briefly revealing the AGN as a traditional quasar, and then eventually shutting down both star formation and black hole growth. After that time, the galaxy exhibited an E$+$A spectrum (a post-starburst galaxy; see Yan et al.\ 2006) and low X-ray luminosities.

Hopkins et al.\ (2008b) showed an optical CMD (their fig.\ 24) featuring AGN and quasars previously presented by Nandra et al.\ (2007) and S\'{a}nchez et al.\ (2004), respectively, and compared the observed host galaxy colours to the colours expected from a merger scenario and from an activation scenario involving secular (that is, in this case, non-merger) processes. They noted that AGN host galaxies typically exhibited colours that placed them on the red sequence or in the upper (redder) region of the blue cloud. Their merger scenario predicts that the AGN host galaxies start in the blue cloud and transition on to the red sequence, while the secular processes are expected to affect the host galaxy colours in the opposite manner. Hopkins et al.\ (2008b) concluded that the colours of observed AGN host galaxies support the merger scenario more strongly than the secular activation scenario.

Croton et al.\ (2006) presented a complementary scenario in which the energy emitted by AGNs with low accretion rates, in high mass galaxies, sufficiently heats gas in the disc to prevent star formation, resulting in rapidly aging stellar populations. We can use observations of AGNs and their host galaxies to test the validity of the predictions suggested by Croton et al.\ (2006), Ciotti \& Ostriker (2007), Hopkins et al.\ (2008a, 2008b) and others. The AGNs typically involved in the simulations are relatively luminous, such as quasars or Seyfert galaxies, many of which should be identifiable by their X-ray emissions. X-ray luminosities can also be used to estimate the growth rate and obscuration of the black holes. Comparing X-ray characteristics to galaxy colours allows a test on the connections and timescales suggested by various models.

As a stellar population ages, its optical colours shift from blue to red, corresponding to a decrease in the temperature and energy output of the stars, so that unbiased galaxy colours indicate the dominant age of the stellar populations in a galaxy. Recent models, such as those described above, make specific predictions about the connection between black hole growth and the ages of stellar populations within a galaxy. The prediction that energetic feedback associated with black hole growth may halt star formation can be tested by comparing observations of galaxy colours to the black hole growth rate and obscuration of the accretion disc around the black hole.

The colours most commonly used to estimate the age of a galaxy's dominant stellar population are optical (e.g.\ $U-B$, $u-r$, $B-V$) and UV-optical (e.g.\ NUV$-R$). Many authors (e.g.\ Baldry et al.\ 2004; Bell et al.\ 2004; Faber et al.\ 2007) have demonstrated that galaxies form a bimodal distribution in a variety of optical colours, and Faber et al.\ (2007) further detected a net flow of galaxies from the `blue cloud' to the `red sequence'. However, young stars contribute significantly to the ultraviolet continuum, suggesting that {\it UV-optical} colours may provide a better indication of {\it recent} star formation (Kennicutt 1998). Wyder et al.\ (2007) studied the UV-optical colour NUV$-R$ and found not only a bimodal distribution but also a significant population between the two colour extremes, now known as the `green valley'. Various studies have since presented observations characterizing the green valley as a possible transition region between the blue cloud and the red sequence (e.g.\ Martin et al.\ 2007; Schiminovich et al.\ 2007), particularly for AGN host galaxies, in which feedback may significantly affect the star formation rates (e.g.\ Schawinski et al.\ 2007; Georgakakis et al.\ 2008b). 

\begin{figure}
\includegraphics[width=3.5in]{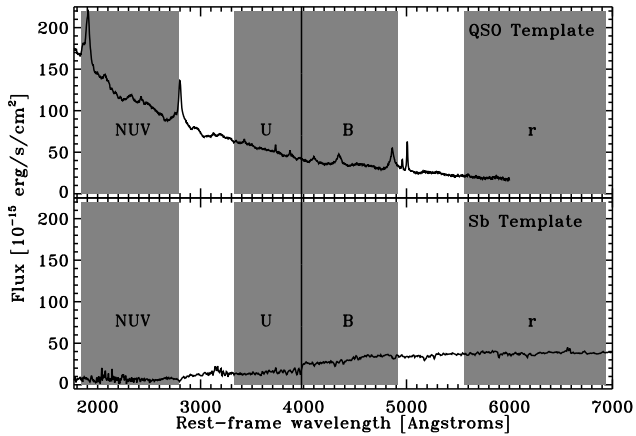}
\caption{Spectral templates from Kinney et al.\ (1996) demonstrating the difference in optical colours between an AGN and a galaxy not hosting an AGN. Approximate wavelength ranges of the NUV, $U$, $B$ and $r$ bands are indicated. The $U$ and $B$ bands do not overlap; the vertical line indicates the separation between them. {\it Upper panel}: QSO spectral template. {\it Lower panel}: Spectral template of an Sb galaxy.}
\label{fig:spectra}
\end{figure}

A full description of an AGN host galaxy depends on measurements of observable features such as colours and the distribution of light in the galaxy. However, the spectrum emitted by an AGN differs significantly from that emitted by its host galaxy and can dominate optical observations. This is most clearly seen in quasi-stellar objects (QSOs), but may also be evident in less luminous AGNs. Fig.\ \ref{fig:spectra} shows this difference in the optical region of the spectrum by showing spectral templates for an AGN (QSO; upper panel) and an Sb galaxy (lower panel) from Kinney et al.\ (1996). Whether considering an optical colour (e.g.\ $U-B$) or a UV-optical colour (e.g.\ NUV$-R$), it is clear that the AGN would appear significantly bluer than the galaxy. The effect on the {\it measured} colour of an AGN host galaxy depends on factors such as the AGN luminosity and obscuration of the AGN accretion disc.

Using a low-redshift ($0.03 < z < 0.07$) sample of optical spectroscopically-selected AGN host galaxies, from which AGNs exhibiting broad spectral lines (typical of the most extreme AGNs, which are most likely to affect optical measurements) were specifically excluded, Kauffmann et al.\ (2007) compared optical colours ($g-r$) from the central regions of the galaxies, based on observations from the Sloan Digital Sky Survey, to integrated (that is, total) UV-optical colours (NUV$-r$) from {\it Galaxy Evolution Explorer}\/ ({\it GALEX}; Martin et al.\ 2005) observations. They showed that light from stars in the {\it outer} regions of these AGN host galaxies dominates the observed UV-optical colours, indicating that the colours are not strongly affected by light from an AGN. However, many current studies that use optical colours to help characterize the host galaxy stellar populations focus on more luminous AGNs (such as those selected by X-ray or radio techniques) at higher redshifts ($z \sim 1$; e.g.\ Nandra et al.\ 2007; Bundy et al.\ 2008; Coil et al.\ 2009; Silverman et al.\ 2009). Thus we also need to understand the possible effects of high-luminosity, $z \sim 1$ AGNs on the measured colours of their host galaxies.

The current work and a companion study (Pierce et al.\ 2010) address this using complementary methods. Pierce et al.\ (2010) added an AGN spectral template (the QSO template shown in Fig.\ \ref{fig:spectra}) to three non-AGN templates (an elliptical galaxy, an Sb galaxy and a starburst galaxy). They scaled the AGN template to contribute a set of specified fractions of the flux from the resulting system and measured the original and resulting optical and UV-optical colours, finding that the AGN template significantly affected the measured colours. The results from Kauffmann et al.\ (2007) could be considered a lower limit and the results from Pierce et al.\ (2010) could be considered an upper limit to the expected effect of an AGN. In addition, to test the potential effect of an AGN on {\it morphology} measurements of the host galaxies, Pierce et al.\ (2010) added a series of optical point sources to optical images of galaxies at $z \sim 0.5$ not known to host AGNs. They compared the measured morphologies of the original and altered galaxy images and found that high AGN fractions can significantly bias the morphology measurements, but that such AGNs are often identifiable from the optical images due to the visibility of the AGN as a central point source. Additional previous work with X-ray-selected AGN at redshifts $0.5 < z < 1.5$, in the Great Observatories Origins Deep Survey (GOODS), found that unobscured AGN hosts are similar to obscured AGN hosts in NUV$-R$, but slightly bluer (Ammons 2009).

The current study was initially undertaken in order to determine the extent to which luminous AGNs at $z \sim 1$ affect the measured colours of their host galaxies. As a result, we have discovered criteria for identifying the AGNs that are most likely to cause colour contamination. Using {\it Hubble Space Telescope}/Advanced Camera for Surveys ({\it HST}/ACS) $V$ and $I$ band images, we measure the outer and nuclear galaxy colours of a sample of X-ray-selected AGNs at redshifts $0.2 < z < 1.2$. Measuring galaxy colours in this manner essentially restricts to the nuclear regions any effect caused by the AGNs, while colours measured for the outer regions are expected to be free of any influence from the AGN. In the special case of QSOs, the outer regions would also be overwhelmed by AGN light, but the sample considered here does not contain any known QSOs.

The AGN host galaxy aperture colours are compared to aperture colours of a control sample consisting predominantly of galaxies not hosting AGNs. We create a control sample because most AGN host galaxies at redshifts $0.2 < z < 1.2$ exhibit characteristics (most importantly, mass and colour) that differ from the characteristics typical of most galaxies at such redshifts. From the AGN host galaxy {\it HST}/ACS images, we determine whether or not the AGN is apparent as an optically visible point source and find that our results correlate with the X-ray obscuration of the AGN, as determined by the X-ray hardness ratio (see Section \ref{data:xray}), indicating a connection between the optical obscuration and the X-ray obscuration. The X-ray hardness ratios are also found to correlate with the outer colours, facilitating comparisons to predictions from the models described above.

The data used for this study come from the All-wavelength Extended Groth Strip International Survey (AEGIS; Davis et al.\ 2007), a multiwavelength survey covering bands from hard X-ray through radio. Many authors have already used these observations to study topics such as galaxy SEDs (Konidaris et al.\ 2007; Symeonidis et al.\ 2007), various aspects of star formation (Ivison et al.\ 2007; Lin et al.\ 2007; Noeske et al.\ 2007a, 2007b; Weiner et al.\ 2007), AGN selection techniques and host galaxy characteristics (Georgakakis et al.\ 2007; Gerke et al.\ 2007; Nandra et al.\ 2007; Pierce et al.\ 2007; Park et al.\ 2008), connections between AGN feedback and galaxy colours (Bundy et al.\ 2008; Georgakakis et al.\ 2008b), galaxy groups and clustering (Fang et al.\ 2007; Georgakakis et al.\ 2008a; Coil et al.\ 2009; Jeltema et al.\ 2009) and a variety of additional topics (Barmby et al.\ 2006; Conselice et al.\ 2007; Huang et al.\ 2007; Kassin et al.\ 2007; Moustakas et al.\ 2007; Wilson et al.\ 2007; Aird et al.\ 2010; Laird et al.\ 2009; Sato et al.\ 2009).

We begin with a description of the data used for the current study (Section \ref{data}) and then describe the selection of AGN and control samples (Section \ref{samples}). Nuclear and outer optical colours and integrated UV-optical colours are presented in Section \ref{results}, followed by a discussion of the results and their implications in Section \ref{discussion}. Finally, in Section \ref{summary} we provide a summary of the main scientific results. Throughout, we use \{$h$, $\Omega_{\Lambda}$, $\Omega_{M}$\} $=$ \{$0.7$, $0.7$, $1-\Omega_{\Lambda}$\} and AB magnitudes, unless otherwise noted. In addition, uncertainties accompanying numerical fractions represent 1$\sigma$ uncertainties, calculated following Gehrels (1986).

%%%%%%%%%%%%%%%%%%%%%%%%%%%%%
% Data
%%%%%%%%%%%%%%%%%%%%%%%%%%%%%
\section{Data}\label{data}

\subsection{Optical images}\label{data:optical}
High spatial resolution {\it HST}/ACS images, with a point spread function (PSF) FWHM of $\sim 0.1$ arcsec, are available for 0.197 deg$^{2}$ of the Extended Groth Strip (EGS). This region was observed in the F606W ($V$) and F814W ($I$) passbands to 5$\sigma$ limiting magnitudes of $V = 28.14$ mag and $I = 27.52$ mag for a point source; the limiting magnitudes are slightly brighter for extended objects (Davis et al.\ 2007).

We measure the observed $V$ and $I$ band light in a series of apertures with radii $0.15$ arcsec, $0.2$ arcsec and $1.5$ arcsec, and then calculate the observed optical colours ($V-I$) within a central region of radius $0.15$ arcsec (the `nuclear' colours) and an annulus having an inner radius of $0.2$ arcsec and an outer radius of $1.5$ arcsec (the `outer' colours). The central region encloses the nuclear point sources that are visibly present in several of the AGN host galaxies, as well as $\sim$80 per cent of the {\it HST}/ACS $V$ band PSF. The outer radius of the annulus is large enough to fully enclose most of the galaxies in our sample, and the size of the inner radius allows a small separation between the outer annulus and the central region. From the observed aperture colours, we estimate the K-corrected, rest-frame aperture colours $(U-B)_{\rm out}$ and $(U-B)_{\rm nuc}$, following methods described by Willmer et al.\ (2006) and Weiner et al.\ (2009). Redshift evolution of the colour gradients presented in Section \ref{results:ub} is minimal and does not significantally influence our results, so that we are not concerned with the differences between the physical scales of the regions measured using the same angular scales at different redshifts.

\begin{figure*}
\includegraphics[width=7in]{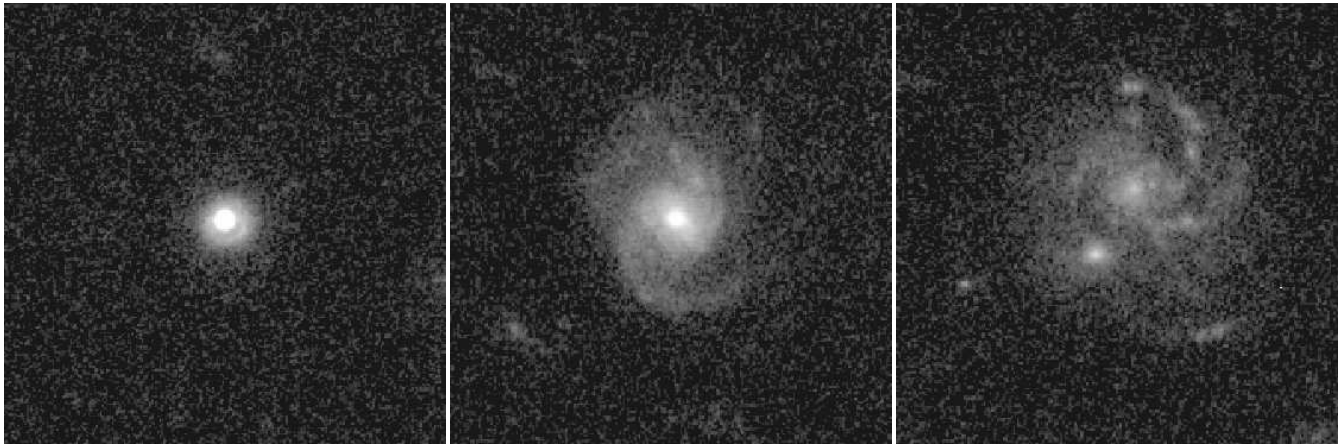}
\caption{AGN host galaxies representing our three point source visibility classifications. From left to right, the galaxies have a `definite point source', a `possible point source' and `no clear point source'. All three $I$ band images were created using the same bias, contrast and logarithmic scale.}
\label{fig:ps_visibility}
\end{figure*}

For each of the X-ray-selected AGNs (Section \ref{samples:agn}), the $V$ and $I$ band images were inspected by-eye to assess the visibility of the AGN as a central (or offset) point source. Criteria used to distinguish between a point source and a region of star formation include size, shape and distinctness of edges; our `point sources' are required to have sizes similar to the ACS PSF (which is independent of redshift), circular shape and distinct edges. Each of the three AGN host galaxies shown in Fig.\ \ref{fig:ps_visibility} represent one of our three point source visibility classifications -- `definite point source', `possible point source' and `no clear point source'. These postage stamps have been created from {\it HST}/ACS $I$ band images, and all use the same logarithmic scale, bias and contrast.

\subsection{Spectroscopic redshifts}\label{data:redshift}
Spectroscopic redshifts are available from the DEEP2 Redshift Survey (Davis et al.\ 2003, 2007) and an MMT Observatory survey of X-ray-selected AGN host galaxies (Coil et al.\ 2009). Redshifts were individually verified and assigned quality codes pertaining to the redshift reliabilities, resulting in $\sim$3700 DEEP2 and 82 MMT `high quality' (reliability confidence levels $\ge 95$ per cent) redshifts in the {\it HST}/ACS-imaged region of the EGS; 80 of these galaxies have high quality spectroscopic redshifts from both the DEEP2 and the MMT surveys.

\subsection{X-ray images}\label{data:xray}
The {\it Chandra X-ray Observatory} ({\it Chandra}) Advanced CCD Imaging Spectrograph observed a strip of eight pointings ($0.67 \deg^{2}$ total) along the EGS for approximately 200 ks per pointing (Nandra et al.\ 2005; Davis et al.\ 2007). Laird et al.\ (2009) presented the AEGIS-X survey (the 1325 X-ray sources detected in the EGS) and described the methods used to reduce and analyse the observations, including calculation of the hardness ratios HR\footnote{HR $\equiv$ (H$-$S)$/$(H$+$S); H$=$2--7 keV counts; S$=$0.5--2 keV counts. HR indicates the amount of attenuation experienced by the lower energy X-rays due to gas and dust in our line-of-sight. At $z=0$, HR $>-0.25$ ($<-0.25$) indicates high (low) attenuation of low-energy X-rays. Due to redshifting of the energy bands, high-$z$ sources may have harder X-ray spectra (more obscuration) than observed.}. For the AEGIS-X survey catalog, X-ray sources are defined as X-ray detections that have at least a 5$\sigma$ detection significance. At $z = 1$, the on-axis flux limit\footnote{The flux limit increases with the distance from the centre of the {\it Chandra} pointing.} for hard-band\footnote{X-ray energy bands: full (0.5--7 keV), soft (0.5--2 keV), hard (2--7 keV) and ultra-hard (4--7 keV)}-selected sources ($f=3.8 \times 10^{-16}$ erg cm$^{-2}$ s$^{-1}$) corresponds to an X-ray luminosity $L_{\rm 2-10 \ keV} = 2.4 \times 10^{42}$ erg s$^{-1}$, slightly in excess of the minimum luminosity used to define our sample of X-ray-selected AGNs (Section \ref{samples:agn}). Thus, although the X-ray AGN sample is complete to $L_{\rm 2-10 \ keV} = 10^{42}$ erg s$^{-1}$ at redshifts $z < 0.7$, it may miss AGNs with $L_{\rm 2-10 \ keV} < 4 \times 10^{42}$ erg s$^{-1}$ at $z = 1.2$ (cf.\ Fig.\ \ref{fig:lx_z}).

\subsection{Stellar masses}\label{data:stellarmass}
Bundy et al.\ (2006) estimated galaxy stellar masses in the four DEEP2 fields by fitting spectral energy distributions (SEDs) based on $B$, $R$, $I$ and $K_{S}$-band images to models created using the Bruzual \& Charlot (2003) stellar population synthesis code. The robustness of these stellar mass estimates for X-ray-selected AGN host galaxies was tested by Bundy et al.\ (2008), and they found that although the fits between the observed and model SEDs are generally {\it better} for galaxies not hosting AGNs, they are typically {\it good} even for AGN host galaxies. The work by Bundy et al.\ (2006) provides stellar mass estimates for 3382 AEGIS galaxies in the {\it HST}/ACS-imaged region, including 54 of our 56 X-ray-selected AGNs (Section \ref{samples:agn}).
% The two galaxies lacking stellar mass estimates from Bundy are 13034617 and 13009690.  Willmer does not provide them either.

Estimates provided by C.\ N.\ A.\ Willmer supplement the stellar mass estimates from Bundy et al.\ (2006). Willmer derived the masses from rest-frame optical $B-V$ colours, following Bell \& de Jong (2001), Bell et al.\ (2005), Willmer et al.\ (2006), Lin et al.\ (2007) and Weiner et al.\ (2009). We test the reliability of these stellar mass estimates for AGN host galaxies by comparing the masses determined from $V-I$ aperture colours using three different inner radii (0.0 arcsec, 0.1 arcsec and 0.2 arcsec) and a common outer radius (1.5 arcsec). The difference in the resulting mass estimates is minimal, indicating that the colour-derived estimates are also robust.

Both stellar mass estimates are available for 390 of our 460 control galaxies (Section \ref{samples:control}), for which the median difference is $\log(M_{*, \rm Bundy}) - \log(M_{*, B-V}) = 0.045$. The masses derived by Bundy et al.\ (2006) typically have lower uncertainties than the masses derived using the $B-V$ colours ($\sim 0.3$ dex). The stellar mass estimates for the AGN hosts (where available) and 96.5 per cent of the control sample galaxies are from Bundy et al.\ (2006); the remaining stellar mass estimates that we use come from Willmer.

\subsection{UV-optical colours}\label{data:nuvr}
Applying the methods described by Salim et al.\ (2005, 2007) to AEGIS galaxies, Salim et al.\ (2009) combined {\it GALEX}\/ and Canada--France--Hawaii Telescope Legacy Survey $u^{*}g'r'i'z'$ observations with $K_{S}$-band photometry to estimate the rest-frame UV-optical colour NUV$-R$, excluding galaxies for which the DEEP2 spectra fit a template for Type-1 (unobscured) AGNs; NUV$-R$ colour estimates are available for 91 per cent (51/56) of our X-ray-selected AGN sample (Section \ref{samples:agn}). Of the five AGNs for which we do not have NUV$-R$ estimates, two exhibit $U-B$ colours and $M_{B}$ magnitudes consistent with Type-1 AGNs (cf.\ fig.\ 1 of Nandra et al.\ 2007) and the remaining three have $U-B$ colours and $M_{B}$ magnitudes consistent with the majority of the AGNs in our sample. These AGNs exhibit approximate X-ray luminosities $3 \times 10^{42}$ erg s$^{-1} < L_{\rm 2-10 \ keV} < 3 \times 10^{44}$ erg s$^{-1}$ and X-ray hardness ratios $-0.59 < $HR $< 0.88$.

Estimating the UV-optical colours includes fitting the galaxy SED to a library of model SED templates and determining the strength of the match (the goodness of fit, $\chi^{2}$; Salim et al.\ 2007). Galaxies for which $\chi^{2} < 10$ are considered to have `reliable' UV-optical colours, and 94 per cent (48/51) of the UV-optical colour estimates for our X-ray sample are thereby deemed reliable. The NUV$-R$ and $V-I$ colours of the three AGN hosts that have `unreliable' UV-optical colours are similar to the colours of the other AGN host galaxies in our sample. Although our results do not depend upon the inclusion or exclusion of these three AGNs, we will present them for comparison along with the galaxies that have reliable colours. Throughout the analyses described here, references to blue, green and red UV-optical colours indicate NUV$-R<3$, $3<$NUV$-R<4.5$ and NUV$-R>4.5$, respectively.

%%%%%%%%%%%%%%%%%%%%%%%%%%%%%
% Samples
%%%%%%%%%%%%%%%%%%%%%%%%%%%%%
\section{Samples}\label{samples}

%%%%%%%%%%%%%%%%%%%%%%%%%%%%%
\subsection{X-ray-selected AGNs at $0.2 < z < 1.2$}\label{samples:agn}
Luminous, high-energy X-ray sources are believed to be AGNs because star formation processes are only expected to account for lower energy and/or less luminous X-ray emissions (e.g.\ Grogin et al.\ 2003, 2005; Barger et al.\ 2005; Laird et al.\ 2005). The current work uses $L_{\rm 2-10 \ keV} > 10^{42}$ erg s$^{-1}$ as the criterion for X-ray-selected AGNs (Grogin et al.\ 2005; Barger, Cowie \& Wang et al.\ 2007), an order of magnitude higher than a conservative cut used by Laird et al.\ (2005) to {\it exclude} AGNs from their sample. Thus, the X-ray-selected AGN sample used here should be fairly pure, though it may exclude some low-luminosity or Compton-thick AGNs.

Georgakakis et al.\ (2009) used the Likelihood Ratio method (e.g.\ Ciliegi et al.\ 2003) to identify DEEP2 counterparts for 131 AEGIS X-ray sources. Using these matches, we follow the method described by Teng et al.\ (2005) to convert the X-ray flux in each of the four X-ray energy bands to 2--10 keV fluxes, assuming a power-law slope $\Gamma = 1.4$ (e.g.\ Peterson 1997), which assumes uniform X-ray obscuration for all of our AGNs. This calculation provides estimates of the observed 2--10 keV X-ray fluxes, which may be less than the {\it intrinsic} X-ray fluxes of the more heavily obscured AGNs. In order to determine the extent to which the choice of $\Gamma$ affects our X-ray selection, we also estimate the X-ray flux and luminosity using a power-law slope appropriate for an unobscured AGNs. Nandra \& Pounds (1994) found a mean intrinsic (unobscured) X-ray spectra with power-law slope $\Gamma =1.95$. For one of our more highly obscured AGNs (HR $ = 0.78$), $\Gamma = 1.4$ indicates an observed luminosity $L_{\rm 2-10 \ keV} = 2.8 \times 10^{42}$ erg s$^{-1}$, while $\Gamma = 2$ suggests an intrinsic luminosity $L_{\rm 2-10 \ keV} = 3.9 \times 10^{42}$ erg s$^{-1}$. Thus, although our calculated luminosities may be lower than the intrinsic X-ray luminosities, the difference is not significant for the results discussed here. In particular, we do not expect to miss more than a few AGNs that, due to heavy obscuration, exhibit observed luminosities below our AGN selection criterion, while having intrinsic X-ray luminosities above the selection criterion.

Hard X-ray luminosities $L_{\rm 2-10 \ keV}$ are then calculated using the extrapolated 2--10 keV fluxes and the spectroscopic redshifts. In order of {\it decreasing} preference, the final X-ray luminosity assigned to an X-ray source is based on the 2--10 keV flux extrapolated from the full, soft, hard, or ultra-hard band flux. The order of preference corresponds to the sensitivity of the bands, such that the most sensitive band (the full band) receives the highest preference; if a particular X-ray source does not have a significant full-band detection, the soft band is tried, and so on.

\begin{figure}
\includegraphics[width=3.5in]{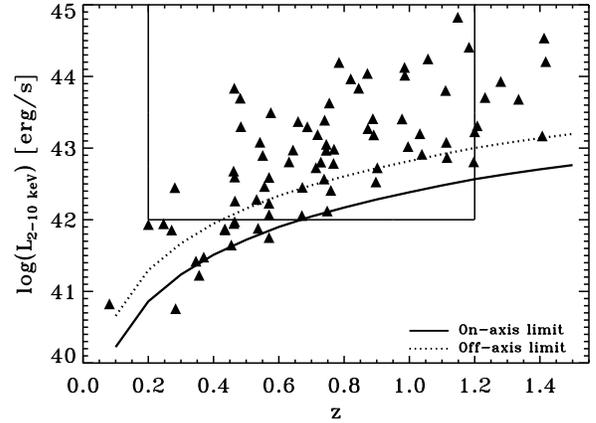}
\caption{X-ray luminosity vs.\ redshift of AGN host galaxies. Luminosities corresponding to the on- and off-axis flux limits of the AEGIS {\it Chandra} 200-ks observations are indicated by the solid and dotted curves, respectively. Horizontal and vertical lines enclose the sample of X-ray-selected AGNs. We expect to miss no more than six objects with X-ray luminosities $L_{\rm 2-10 \ keV} < 10^{42.5}$ erg s$^{-1}$ at $0.7 < z < 1.2$.}
\label{fig:lx_z}
\end{figure}

Fig.\ \ref{fig:lx_z} shows the 2--10 keV luminosity as a function of redshift for X-ray sources with redshifts $z < 1.5$, represented by solid black triangles. Also plotted are luminosities corresponding to the on-axis (solid curve) and off-axis (dotted curve) flux limits for the 200-ks AEGIS {\it Chandra} observations. The horizontal and vertical lines at $L_{\rm 2-10 \ keV} = 10^{42}$ erg s$^{-1}$, $z = 0.2$ and $z = 1.2$ enclose the 56 X-ray-selected AGNs discussed in the current work. Note that the survey flux limits at redshifts $z > 0.7$ exceed the minimum X-ray luminosity requirement for the AGN selection used here; as a result of this incompleteness, our AGN sample is estimated to be missing up to six objects with X-ray luminosities $L_{\rm 2-10 \ keV} < 10^{42.5}$ erg s$^{-1}$ at $0.7 < z < 1.2$. Table \ref{table:xray_sample} lists several relevant measurements of our AGNs and their host galaxies.

\begin{table*}
\centering
\begin{minipage}{175mm}
\caption{Description of X-ray sample. OBJNO refers to the object number in the DEEP2 catalog.\label{table:xray_sample}}
\begin{tabular}{@{}cccccccccccc@{}}
OBJNO & z & $\log(L_{\rm X})$ & $M_B$ & $M_R$ & $\log(M_{*})$ & NUV$-R$ & $U-B$ & $(U-B)_{\rm out}$ & $(U-B)_{\rm nuc}$ & HR & Point Source? \\
   &   & [erg s$^{-1}$]             & [AB]  & [AB]  & [$\Msun$]     & [AB]    & [AB]  & [AB]              & [AB]              &    & \\
\hline
12004450 & 0.281 & 42.45 & -20.51 & -99.00 &  10.91 & -99.00 & 1.164 & 1.254 & 1.075 &  0.782 & No \\
12007878 & 0.985 & 44.13 & -22.29 & -23.30 &  11.34 &   4.73 & 1.301 & 1.252 & 1.273 &  0.086 & Maybe \\
12007896 & 0.671 & 42.06 & -21.44 & -22.28 &  10.76 &   2.46 & 0.680 & 0.687 & 0.765 & -0.074 & Maybe \\
12007926 & 0.873 & 43.27 & -21.78 & -22.87 &  11.14 &   4.01 & 1.066 & 1.090 & 0.990 &  0.717 & Maybe \\
12007954 & 1.148 & 44.83 & -24.07 & -24.81 &  11.93 &   3.92 & 1.185 & 1.218 & 1.292 & -0.182 & Yes \\
12007962 & 0.996 & 43.02 & -22.11 & -23.37 &  11.50 &   5.31 & 1.306 & 1.313 & 1.326 &  0.373 & No \\
12008225 & 0.482 & 43.70 & -21.77 & -99.00 &  10.55 & -99.00 & 0.182 & 0.503 & 0.045 & -0.310 & Yes \\
12008608 & 0.532 & 42.28 & -21.42 & -22.74 &  11.35 &   4.11 & 1.173 & 1.194 & 1.111 &  0.031 & Maybe \\
12012431 & 0.902 & 42.73 & -21.05 & -22.16 &  10.77 &   3.11 & 0.994 & 0.865 & 0.831 & -0.619 & No \\
12012467 & 0.986 & 44.02 & -22.05 & -23.28 &  11.34 &   4.94 & 1.167 & 1.195 & 0.922 & -0.244 & Maybe \\
12012471 & 0.484 & 43.30 & -20.47 & -21.30 &  10.15 &   1.94 & 0.428 & 0.775 & 0.205 & -0.617 & Yes \\
12012474 & 0.465 & 42.60 & -20.24 & -21.37 &  10.55 &   3.53 & 0.871 & 1.021 & 0.988 & -0.443 & Yes \\
12012543 & 0.551 & 42.90 & -20.93 & -21.87 &  11.43 &   2.59 & 0.750 & 0.716 & 1.179 &  0.021 & No \\
12015703 & 1.038 & 42.91 & -21.93 & -22.70 &  11.12 &   4.33 & 1.244 & 1.236 & 1.265 &  0.856 & No \\
12016316 & 0.719 & 43.19 & -20.81 & -21.60 &  10.66 &   2.62 & 0.855 & 0.858 & 1.000 &  0.774 & No \\
12019829 & 0.570 & 42.07 & -19.72 & -20.99 &  10.53 &   4.24 & 1.191 & 1.081 & 0.967 &  0.118 & No \\
12020028 & 0.570 & 42.23 & -22.13 & -23.52 &  11.58 &   6.04 & 1.344 & 1.298 & 1.219 & -0.486 & Maybe \\
12020452 & 1.057 & 44.25 & -22.11 & -22.85 &  10.40 &   1.81 & 0.533 & 0.465 & 0.676 &  0.014 & Yes \\
12024136 & 0.729 & 42.81 & -21.94 & -22.71 &  10.91 &   2.80 & 0.784 & 0.797 & 0.559 &  0.018 & Maybe \\
12024309 & 0.643 & 42.97 & -21.73 & -22.64 &  10.96 &   2.63 & 0.810 & 0.810 & 0.651 & -0.348 & Maybe \\
12024323 & 0.570 & 42.59 & -20.92 & -22.19 &  10.96 &   3.51 & 1.152 & 1.104 & 1.213 & -0.028 & Maybe \\
12024913 & 0.759 & 42.41 & -21.35 & -22.05 &  10.52 &   2.11 & 0.583 & 0.526 & 0.657 & -0.624 & No \\
12025302 & 0.464 & 43.84 & -19.71 & -20.93 &  10.52 &   3.26 & 1.021 & 0.994 & 1.290 &  0.465 & Maybe \\
12027585 & 0.784 & 44.20 & -21.48 & -22.54 &  10.92 &   2.97 & 0.777 & 0.817 & 0.446 & -0.371 & Yes \\
12028330 & 0.820 & 43.97 & -21.83 & -22.93 &  11.15 &   4.07 & 1.170 & 1.084 & 1.242 & -0.196 & No \\
12028367 & 0.845 & 43.84 & -21.86 & -22.75 &  11.13 &   2.89 & 0.760 & 0.928 & 0.071 & -0.429 & Yes \\
12028639 & 0.977 & 43.41 & -21.11 & -22.44 &  10.99 &   3.79 & 1.007 & 0.863 & 0.594 & -0.657 & No \\
13004291 & 1.197 & 42.81 & -22.62 & -23.71 &  11.52 &   2.94 & 0.765 & 0.736 & 0.677 & -0.343 & Maybe \\
13010503 & 0.461 & 42.68 & -19.58 & -20.76 &  10.40 &   4.22 & 1.023 & 0.983 & 0.750 & -0.330 & Maybe \\
13011701 & 0.740 & 43.39 & -21.72 & -22.52 &  11.09 &   3.33 & 0.842 & 0.830 & 0.575 & -0.425 & Maybe \\
13018886 & 1.032 & 43.20 & -21.22 & -21.87 &  10.51 &   2.65 & 0.850 & 0.832 & 0.914 & -0.509 & No \\
13018972 & 0.659 & 43.37 & -19.24 & -20.38 &  10.10 &   4.20 & 1.145 & 1.008 & 1.057 & -0.167 & No \\
13019240 & 0.745 & 43.05 & -21.04 & -22.09 &  10.80 &   5.30 & 1.150 & 1.103 & 1.088 &  0.154 & No \\
13019950 & 0.769 & 42.98 & -21.71 & -22.83 &  11.20 &   4.30 & 1.073 & 1.098 & 0.956 &  0.322 & Maybe \\
13025417 & 0.745 & 42.97 & -21.68 & -22.89 &  11.10 &   4.02 & 1.224 & 1.232 & 1.258 & -0.592 & No \\
13025494 & 0.739 & 42.57 & -21.29 & -22.49 &  11.10 &   5.60 & 1.233 & 1.130 & 1.277 & -0.200 & No \\
13026080 & 0.672 & 42.45 & -18.91 & -19.86 &   9.83 &   2.82 & 0.900 & 0.916 & 0.598 & -0.479 & No \\
13026185 & 0.631 & 42.81 & -20.78 & -22.02 &  10.87 &   4.71 & 1.173 & 1.105 & 0.922 & -0.170 & Maybe \\
13026574 & 1.113 & 43.08 & -21.91 & -22.90 &  11.14 &   4.43 & 1.090 & 1.160 & 1.260 & -0.159 & No \\
13027442 & 0.768 & 42.79 & -22.06 & -23.25 &  11.24 &   4.34 & 1.292 & 1.202 & 1.284 & -0.354 & No \\
13034447 & 0.748 & 42.12 & -21.61 & -22.66 &  11.01 &   4.54 & 1.139 & 1.100 & 1.252 &  0.040 & Maybe \\
13034617 & 1.111 & 43.80 & -22.08 & -23.44 &  10.70 &   3.29 & 0.889 & 0.854 & 0.436 & -0.443 & Yes \\
13035123 & 1.115 & 42.87 & -23.84 & -24.24 &  11.53 &   2.12 & 0.892 & 0.843 & 0.859 & -0.257 & No \\
13040909 & 1.183 & 44.41 & -23.20 & -99.00 &  10.73 & -99.00 & 0.398 & 0.472 & 0.047 & -0.305 & Yes \\
13042206 & 0.891 & 43.19 & -21.53 & -22.52 &  10.98 &   4.23 & 1.217 & 1.134 & 1.084 &  0.243 & No \\
13042378 & 0.897 & 42.53 & -20.77 & -21.69 &  10.62 &   3.65 & 1.141 & 1.139 & 1.211 &  0.090 & No \\
13042389 & 0.556 & 42.46 & -18.48 & -20.13 &  10.51 &   3.75 & 0.707 & 1.055 & 0.615 &  0.176 & No \\
13042603 & 0.888 & 43.41 & -21.09 & -22.46 &  10.90 &   3.02 & 0.828 & 0.801 & 0.340 & -0.239 & No \\
13049115 & 0.465 & 42.26 & -20.93 & -21.46 &  10.48 &   2.01 & 0.670 & 0.615 & 0.531 &  0.008 & No \\
13050479 & 0.687 & 43.30 & -20.91 & -22.12 &  11.07 &   4.36 & 1.246 & 1.109 & 1.215 &  0.611 & No \\
13058137 & 1.200 & 43.23 & -22.09 & -99.00 &  10.97 & -99.00 & 0.992 & 0.889 & 0.600 & -0.594 & No \\
13058235 & 0.714 & 42.73 & -20.77 & -21.79 &  10.76 &   4.28 & 1.194 & 1.099 & 1.115 &  0.844 & Maybe \\
13101998 & 0.755 & 43.63 & -21.34 & -22.39 &  10.64 &   2.56 & 0.776 & 0.791 & 0.443 & -0.289 & Yes \\
13018061 & 0.575 & 43.49 & -21.61 & -22.66 &  10.50 &   3.25 & 0.776 & 0.855 & 0.507 & -0.306 & Yes \\
13026061 & 0.871 & 44.04 & -20.91 & -21.24 &  10.62 &   1.43 & 0.635 & 0.861 & 0.020 & -0.349 & Maybe \\
13009690 & 0.542 & 43.08 & -20.13 & -99.00 & -99.00 & -99.00 & 1.167 & 1.139 & 1.052 &  0.880 & Yes \\
\hline
\end{tabular}
\end{minipage}
\end{table*}

%%%%%%%%%%%%%%%%%%%%%%%%%%%%%
\subsection{Control sample}\label{samples:control}
Intrinsic properties (such as stellar masses and colours) of X-ray-selected AGN host galaxies differ from the general galaxy population (e.g.\ Kauffmann et al.\ 2003; Lacy et al.\ 2004; Hatziminaoglou et al.\ 2005; Stern et al.\ 2005; Barmby et al.\ 2006; Nandra et al.\ 2007; Bundy et al.\ 2008). Comparing the colour gradients of AGN host galaxies to the colour gradients of quiescent galaxies occupying, for example, a different redshift range, may significantly affect the results inferred from the comparison. To guard against this, we create a control sample designed to match the AGN sample in redshift range and stellar mass range. For each X-ray-selected AGNs with redshift $z_{\rm AGN}$ and stellar mass $M_{*,\rm AGN}$, we include in the control sample all of the galaxies that have redshifts $z_{\rm galaxy}$ and stellar masses $M_{*,\rm galaxy}$ such that
\begin{equation}
\frac{|z_{\rm AGN} - z_{\rm galaxy}|}{z_{\rm max} - z_{\rm min}} < 0.03
\end{equation}
and
\begin{equation}
\frac{|\log(M_{*,\rm AGN}) - \log(M_{*,\rm galaxy})|}{\log(M_{*,\rm max}) - \log(M_{*,\rm min})} < 0.03,
\end{equation}
where $z_{\rm min}$, $z_{\rm max}$, etc., are the minimum and maximum redshifts and stellar mass estimates for the AGN sample. For completeness, the control sample includes our AGN host galaxies (12 per cent of the control sample galaxies are identified as AGNs). The value 0.03 helps to minimize the effect of cosmic variance on the number of control galaxies assigned to each AGN host galaxy. One of the 56 AGN host galaxies does not have an available stellar mass, so there are no control galaxies specifically matched to this system. However, it has X-ray luminosities, redshifts, $U-B$ colours and $M_{B}$ magnitudes typical of the AGN sample.

\begin{figure}
\includegraphics[width=3.5in]{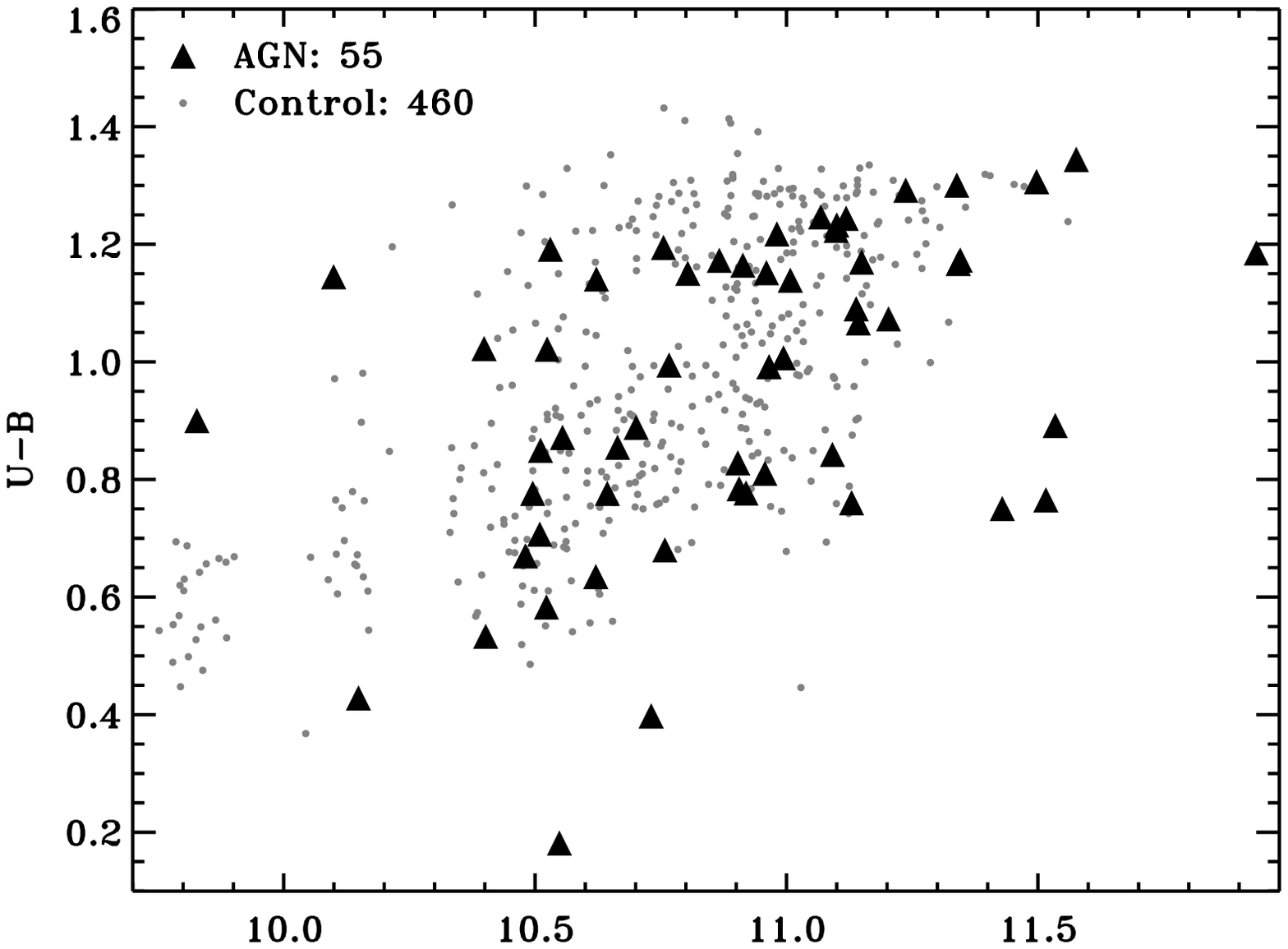}
\includegraphics[width=3.5in]{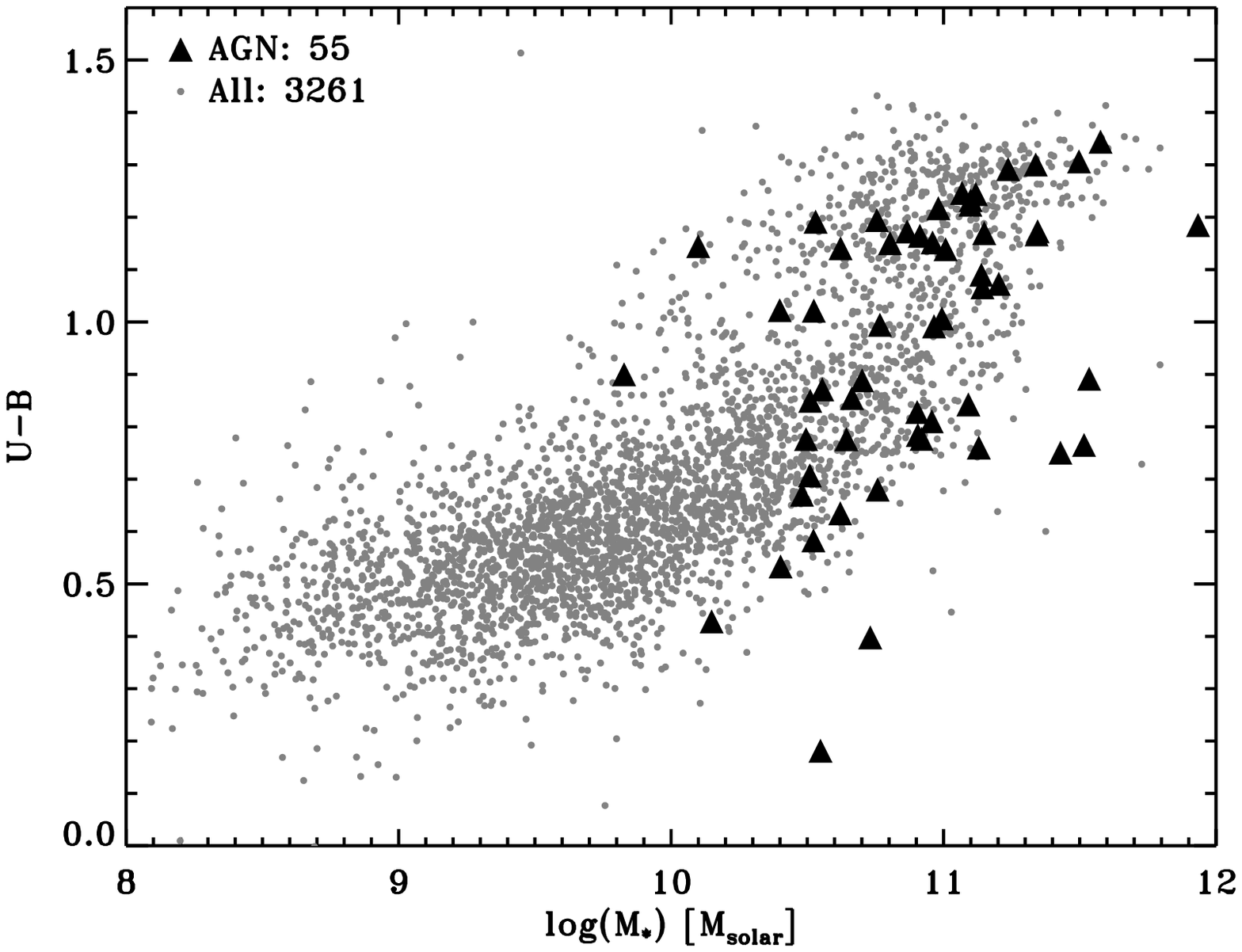}
\caption{Optical colour vs.\ stellar mass. Triangles represent AGN host galaxies; small gray circles represent the corresponding control galaxies ({\it upper panel}) or all galaxies at redshift $0.2<z<1.2$ for which we have stellar masses ({\it lower panel}). The number of objects in each sub-sample is indicated on the figure. Note that the axes change scale between the two panels to accomodate the samples presented. On account of the approximate correlation between colour and stellar mass we do not explicitly use galaxy colour when selecting galaxies for the control sample.}
\label{fig:ub_mstar}
\end{figure}

Typical AGN host galaxies exhibit redder colours and higher stellar masses than the majority of galaxies not hosting AGNs (e.g.\ Lacy et al.\ 2004; Hatziminaoglou et al.\ 2005; Nandra et al.\ 2007; Schawinski et al.\ 2007), suggesting that our control sample should approximately match the AGN sample with respect to galaxy colour. However, we do not find it necessary to explicitly match the colours of the AGN host galaxies to those of the control sample because of an approximate correlation between galaxy colour and estimated stellar mass, as shown in Fig.\ \ref{fig:ub_mstar}. Among the least massive AGN host galaxies, the approximation of a correlation between colour and stellar mass appears to break down, but for the majority of the AGN host galaxies, it seems reasonable to select the control sample only on the basis of redshift and stellar mass.

\begin{figure}
\includegraphics[width=3.5in]{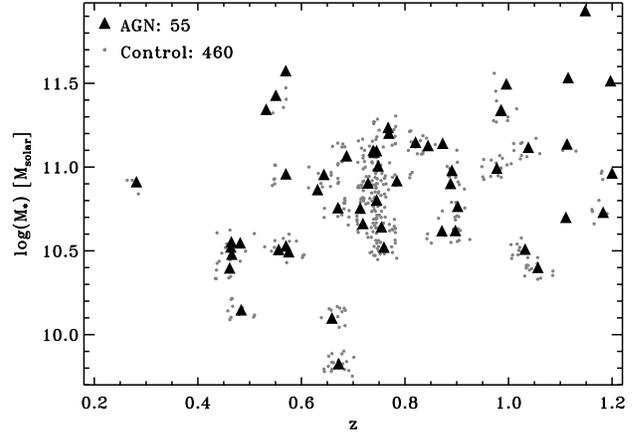}
\caption{Stellar mass vs.\ redshift. Triangles represent AGN host galaxies; small gray circles represent the corresponding control sample. The number of objects in each sub-sample is indicated on the figure. AGNs are included in the control sample.}
\label{fig:mstar_z}
\end{figure}

Fig.\ \ref{fig:mstar_z} shows the stellar masses and redshifts of the AGN and control sample galaxies. As mentioned previously, one of the 56 AGN host galaxies is not represented on this figure, due to an unavailable stellar mass estimate. Fifty-four of the remaining 55 AGN host galaxies span a redshift range $0.4 < z < 1.2$, and 54 have estimated stellar masses $10 < \log(M_{*}/\Msun) < 12$. Due to the redshift distribution of the underlying AEGIS galaxies, which peaks at $z \sim 0.75$ (cf.\ fig.\ 2 of Wilson et al.\ 2007), AGN hosts with redshifts $0.7 < z < 0.8$ have a relative overabundance of corresponding control galaxies (14 control galaxies/AGN host galaxy), while hosts with redshifts $0.2 < z < 0.7$ or $0.8 < z < 1.2$ have a relative underabundance ($\sim 5$ control galaxies/AGN host galaxy).

%%%%%%%%%%%%%%%%%%%%%%%%%%%%%
\subsection{Colour-magnitude diagrams of the AGN and control samples}\label{samples:cmds}
\begin{figure}
\includegraphics[width=3.5in]{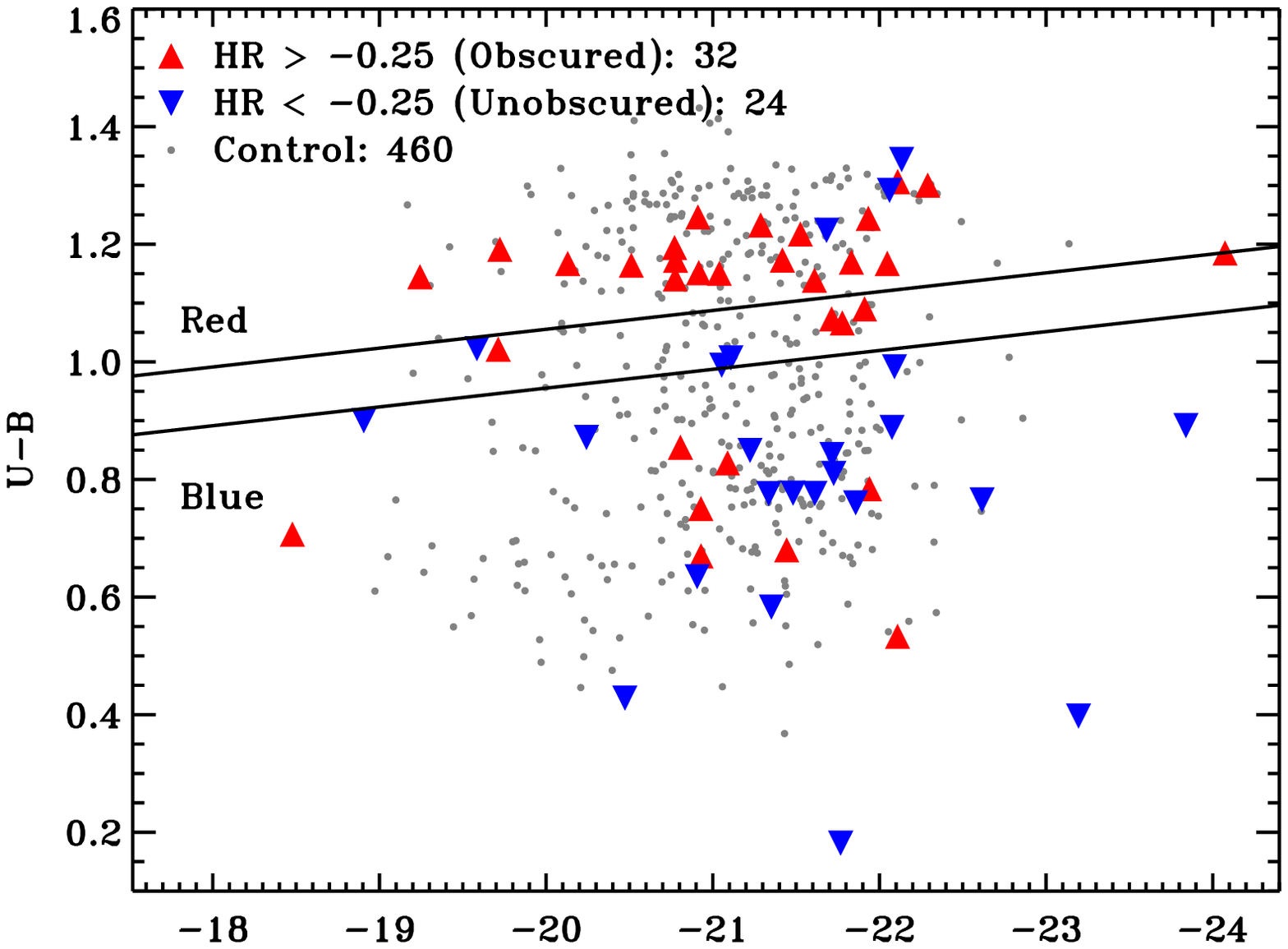}
\includegraphics[width=3.5in]{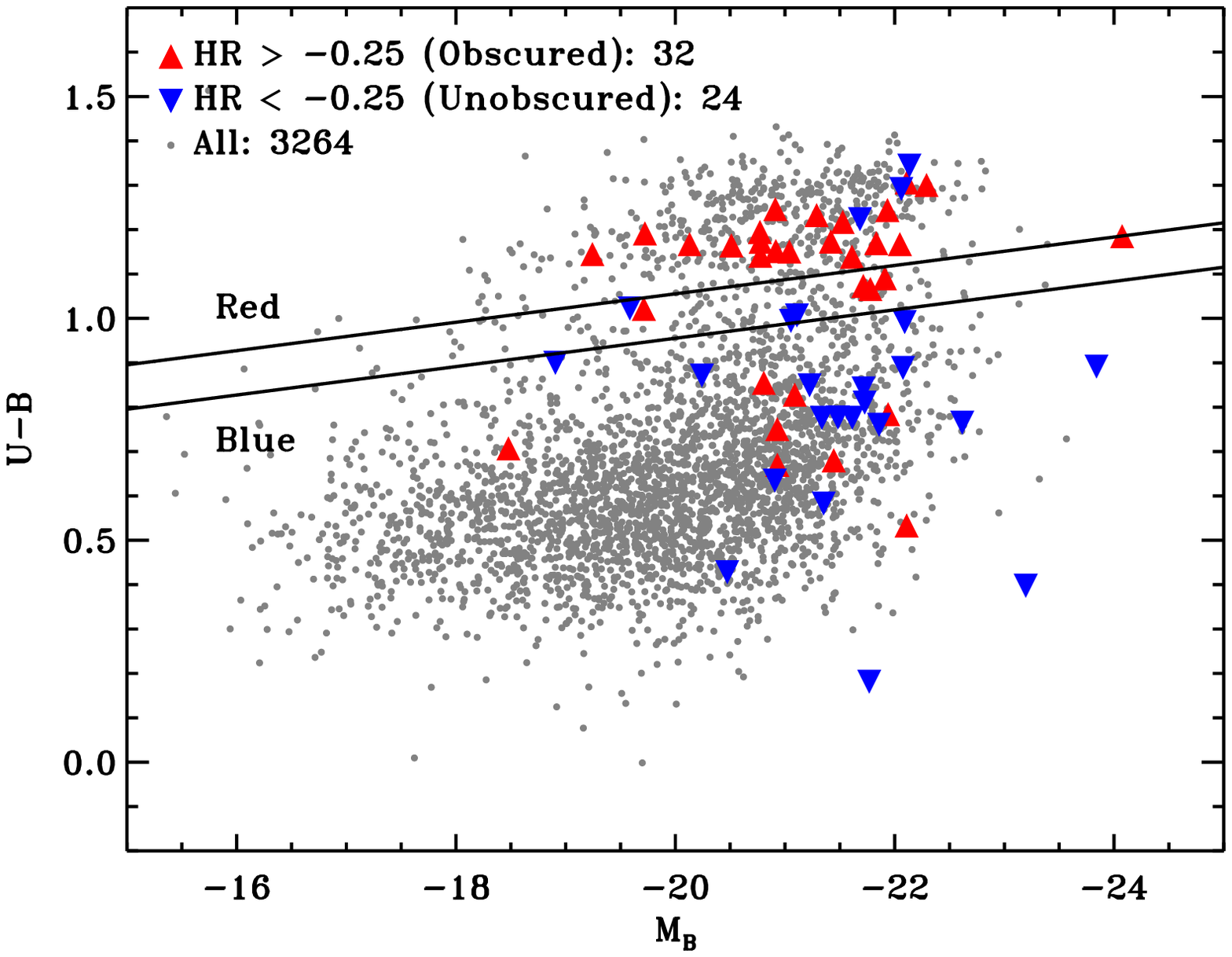}
\caption{Optical colour-magnitude diagrams. Triangles represent AGN host galaxies; small gray circles represent the corresponding control sample ({\it upper panel}) or all galaxies with redshifts $0.2<z<1.2$ ({\it lower panel}). Triangle colour and orientation indicate the level of X-ray obscuration, as noted on the plot. Diagonal lines separate the CMD into the optical blue cloud and red sequence. Note that the axes change scale between the two panels to best accomodate the sub-samples presented. The five AGN outliers are described in the text.}
\label{fig:ub_mb}
\end{figure}

In Fig.\ \ref{fig:ub_mb} we present the optical colour-magnitude diagram (CMD) for the AGN and control samples. Most (52/56) of our AGN host galaxies exhibit colours and $M_{B}$ magnitudes consistent with those of the control sample. The four outliers, which we describe below (see Table \ref{table:xray_sample} for additional details), are generally brighter and/or bluer than galaxies in the control sample.

\begin{enumerate}
\item OBJNO 12007954 [($M_{B}$, $U-B$) $=$ (-24.07, 1.19)]: Le Floc'h et al.\ (2007) discussed this hyper-luminous infrared galaxy (hyper-LIRG; $L_{IR} > 10^{12} \Lsun$), which has a foreground galaxy that may affect the observed photometry. This galaxy has a UV-optical colour NUV$-R = 3.92$, $M_{R} = -24.81$ and X-ray hardness ratio HR $ = -0.182$.

\item OBJNO 13035123 [($M_{B}$, $U-B$) $=$ (-23.84, 0.89)]: This X-ray-selected AGN host galaxy lacks optical spectral features typical of AGNs. It is the third galaxy shown in Fig.\ \ref{fig:ps_visibility}, exhibiting low surface brightness and strong post-starburst characteristics; dust reddening may affect the observed colours, and the UV-optical colour is available but considered unreliable.

\item OBJNO 13040909 [($M_{B}$, $U-B$) $=$ (-23.20, 0.40)]: This AGN host galaxy is the first galaxy shown in Fig.\ \ref{fig:ps_visibility}. Its bright blue point source may affect the reliability of the measured colours and brightness, particulary because the K-corrections were derived using a galaxy SED model, which may not properly reflect the QSO-like spectrum (i.e.\ weak broad lines) of this object.

\item OBJNO 12008225 [($M_{B}$, $U-B$) $=$ (-21.77, 0.18)]: The final outlier exhibits broad lines in its optical spectrum and features a bright blue point source, which may again affect the estimated K-correction. Therefore, the apparent colour and brightness may be lower than the intrinsic values.
\end{enumerate}

The control sample shown in Fig.\ \ref{fig:ub_mb} includes many faint blue galaxies (i.e.\ $M_{B} > -20.8$ and $U-B < 0.8$) that lack obvious AGN counterparts. However, they are within the {\it ranges} of the AGN host galaxy colours and $B$ band magnitudes. Almost half of these galaxies correspond to the least massive AGN host galaxy in our sample (cf.\ Fig.\ \ref{fig:mstar_z}), and all but one correspond to AGN host galaxies less massive than $\log(M_{*}/\Msun) < 10.6$. The inclusion of these galaxies does not seem to affect the general results presented in later sections.

Nandra et al.\ (2007) also presented a CMD featuring AEGIS X-ray sources and non-sources, and in order to properly compare the two figures, several differences between them need to be understood. First note the following differences regarding the X-ray sample: (1) Nandra et al.\ (2007) required spectroscopic redshifts $0.6 < z < 1.4$ instead of our requirement of $0.2 < z < 1.2$; and (2) our sample is restricted to the AEGIS region imaged by {\it HST}/ACS, while Nandra et al.\ (2007) also included X-ray sources external to the ACS-imaged region. Together, these two differences explain the absence in Fig.\ \ref{fig:ub_mb} of five bright blue AGN hosts that are shown by Nandra et al.\ (2007); these objects have redshifts $1.2 < z < 1.4$ and/or are located outside of the ACS-imaged region. With respect to the galaxy samples, note that Nandra et al.\ (2007) included the faint blue-cloud galaxies that are generally not selected as part of our control sample (but are shown in the lower panel of Fig.\ \ref{fig:ub_mb}). With respect to the plots themselves, the vertical axis of the upper panel of our Fig.\ \ref{fig:ub_mb} differs from the CMD shown by Nandra et al.\ (2007) in order to better present our sample of mostly red-sequence galaxies.

\begin{figure}
\includegraphics[width=3.5in]{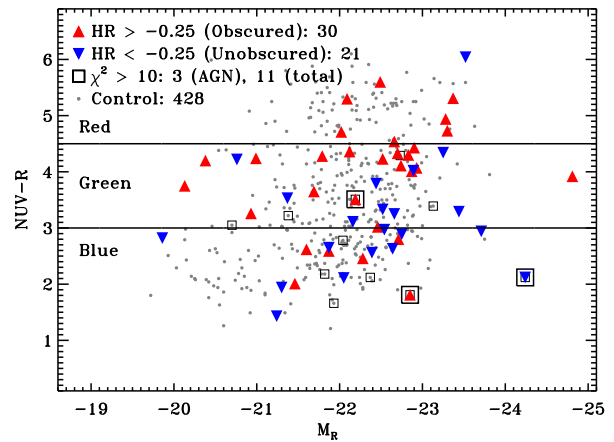}
\caption{UV-optical colour-magnitude diagram. Symbols are as in Fig.\ \ref{fig:ub_mb}, except that boxed symbols represent galaxies for which $\chi^{2} > 10$, indicating unreliable UV-optical colours; horizontal lines separate the CMD into the UV-optical blue cloud, green valley and red sequence. The reliability of UV-optical colours appears to be colour-dependent; unreliable UV-optical colours are rare among red-sequence galaxies.}
\label{fig:nuvr_mr}
\end{figure}

%%% NUV-R CMD %%%
Fig.\ \ref{fig:nuvr_mr} shows the {\it UV-optical} CMD of the 428 control galaxies (51 of which host AGNs) for which we have NUV$-R$ colour estimates. Boxed symbols indicate the 11 control galaxies (three of which host AGNs) for which the NUV$-R$ colour estimates are considered unreliable ($\chi^{2} > 10$; cf.\ Section \ref{data:nuvr}). From the AGN sample we have UV-optical colours for 91 per cent of the galaxies and {\it reliable} UV-optical colours for 86 per cent. In comparison, we have UV-optical colours for 93 per cent of the control sample and {\it reliable} UV-optical colours for 91 per cent. For analyses involving UV-optical colours, we only include the 417 control galaxies (including 48 AGN host galaxies) for which we have reliable UV-optical colours.

Potentially unreliable UV-optical colours appear to be more common among green-valley and blue-cloud galaxies than among red-sequence galaxies. The control galaxies that have unreliable UV-optical colours do not otherwise differ significantly from the general control sample. We do not attempt to calculate the UV-optical colours of AGN hosts for which the corresponding DEEP2 spectra fit a Type-1 (unobscured) template (Section \ref{data:nuvr}). Such excluded AGN hosts include two of the lower luminosity X-ray-obscured systems (OBJNOs 12004450, 13009690), one of the lower luminosity X-ray-unobscured systems (OBJNO 13058137) and two of the higher luminosity X-ray-unobscured systems (OBJNOs 12008225, 13040909). The nuclei of all five excluded AGN hosts are at least slightly bluer than the outer regions, but we do have reliable UV-optical colours for many similar systems. In contrast, the nuclei of the three AGN hosts (OBJNOs 12020452, 12024323, 13035123) for which NUV$-R$ colours are available but considered unreliable are at least slightly redder than their outer colours. Because of the varied characteristics of the excluded AGN host galaxies, the sub-samples for which we have reliable UV-optical colours are still representative of the original samples.

%%%%%%%%%%%%%%%%%%%%%%%%%%%%%
% Results
%%%%%%%%%%%%%%%%%%%%%%%%%%%%%
\section{Results}\label{results}

%%%%%%%%%%%%%%%%%%%%%%%%%%%%%
\subsection{$U-B$ aperture colours}\label{results:ub}
\begin{figure}
\includegraphics[width=3.5in]{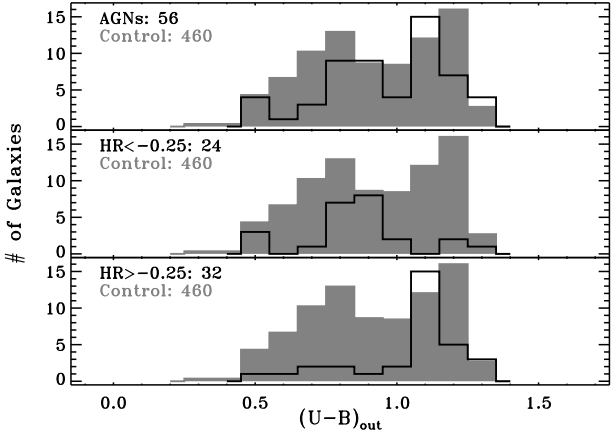}
\caption{Distributions of the outer colour $(U-B)_{\rm out}$ (measured using annuli with radii 0.2 arcsec and 1.5 arcsec) for the AGN and control samples. The number of control sample galaxies is normalized for comparison with the AGN sample. {\it Top panel}: black open histogram represents the full AGN sample; gray solid histogram represents the control sample. {\it Middle panel}: black open histogram represents unobscured AGNs; gray solid histogram represents the control sample. {\it Bottom panel}: black open histogram represents obscured AGNs; gray solid histogram represents the control sample. The AGN and control samples both exhibit colour bimodality, but the outer colours of red AGN hosts are bluer than those of the control galaxies, while the outer colours of blue AGN hosts are redder than those of the control galaxies. Unexpectedly, in the outer regions of the galaxies X-ray-unobscured AGNs exhibit blue colours, while X-ray-obscured AGNs exhibit red colours.}
\label{fig:hist_ub_out}
\end{figure}

Fig.\ \ref{fig:hist_ub_out} presents the distribution of the {\it outer} $U-B$ colours, measured within an annulus having an inner radius at $0.2$ arcsec and an outer radius at $1.5$ arcsec, for the AGN sample (open histograms) and the control sample (solid histograms). Though both the AGN and control samples exhibit red-blue colour bimodalities, the {\it red} AGN host galaxies tend to be bluer than the red control galaxies, while the {\it blue} AGN host galaxies are typically redder than the blue control galaxies. Because most of the colours are not likely to be contaminated by blue AGN light, the AGN host galaxies appear to have younger stellar populations than typical red-sequence galaxies.

A two-dimensional Kolmogorov--Smirnov test (K--S test; Fasano \& Franceschini 1987) indicates that the outer colours of the control and AGN samples exhibit only a 15 per cent probability of having been drawn from the same parent population. By splitting the AGN sample into X-ray-unobscured (HR $< -0.25$) and X-ray-obscured (HR $> -0.25$) sub-samples, as shown in the middle and bottom panels, respectively, we find that X-ray-unobscured AGNs comprise most of the blue peak, while most AGNs comprising the red peak are X-ray-obscured. The outer colours of the obscured and unobscured samples show a very low probability ($<<1$) of having been drawn from the same parent population and probabilities of 0.05 per cent and 4 per cent, respectively, of having been drawn from the same parent population as the control sample.

\begin{figure}
\includegraphics[width=3.5in]{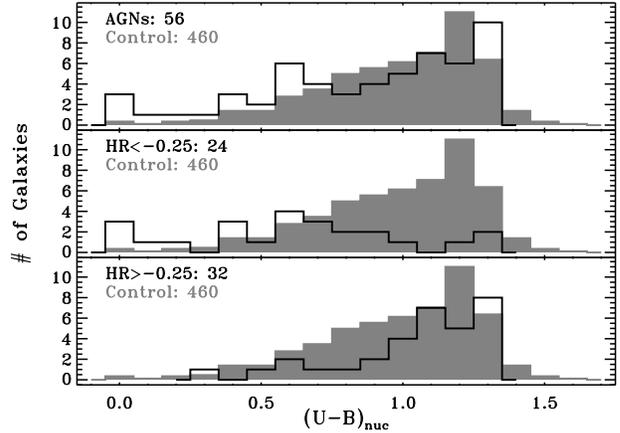}
\caption{Distributions of the nuclear colour $(U-B)_{\rm nuc}$ (measured using circles of radius 0.15 arcsec) for the AGN and control samples. The number of control sample galaxies is normalized to faciliate comparison with the AGN sample. Panels and histograms are as in Fig.\ \ref{fig:hist_ub_out}. Although AGN hosts still show a colour bimodality, the control galaxies have a single red peak.}
\label{fig:hist_ub_nuc}
\end{figure}

{\it Nuclear} $U-B$ colours (from the region enclosed by a circle of radius $0.15$ arcsec) are shown in Fig.\ \ref{fig:hist_ub_nuc}; the top panel again presents the control sample and the entire AGN sample. The nuclear colours of the control sample galaxies are not bimodal; instead the distribution features a red peak and a blue tail. The AGN colour distribution also features a red peak, though with a more substantial blue tail. For the nuclear colours, a K--S test indicates a low probability (2 per cent) that the control and AGN samples were drawn from the same parent population. The lower two panels of Fig.\ \ref{fig:hist_ub_nuc} show that AGN host galaxy nuclear colours are also correlated with X-ray obscuration. The predominant red peak of the AGN colour distribution mostly includes X-ray-obscured AGNs (lower panel), while the blue tail is mainly comprised of X-ray-unobscured AGNs (middle panel).

\begin{figure}
\includegraphics[width=3.5in]{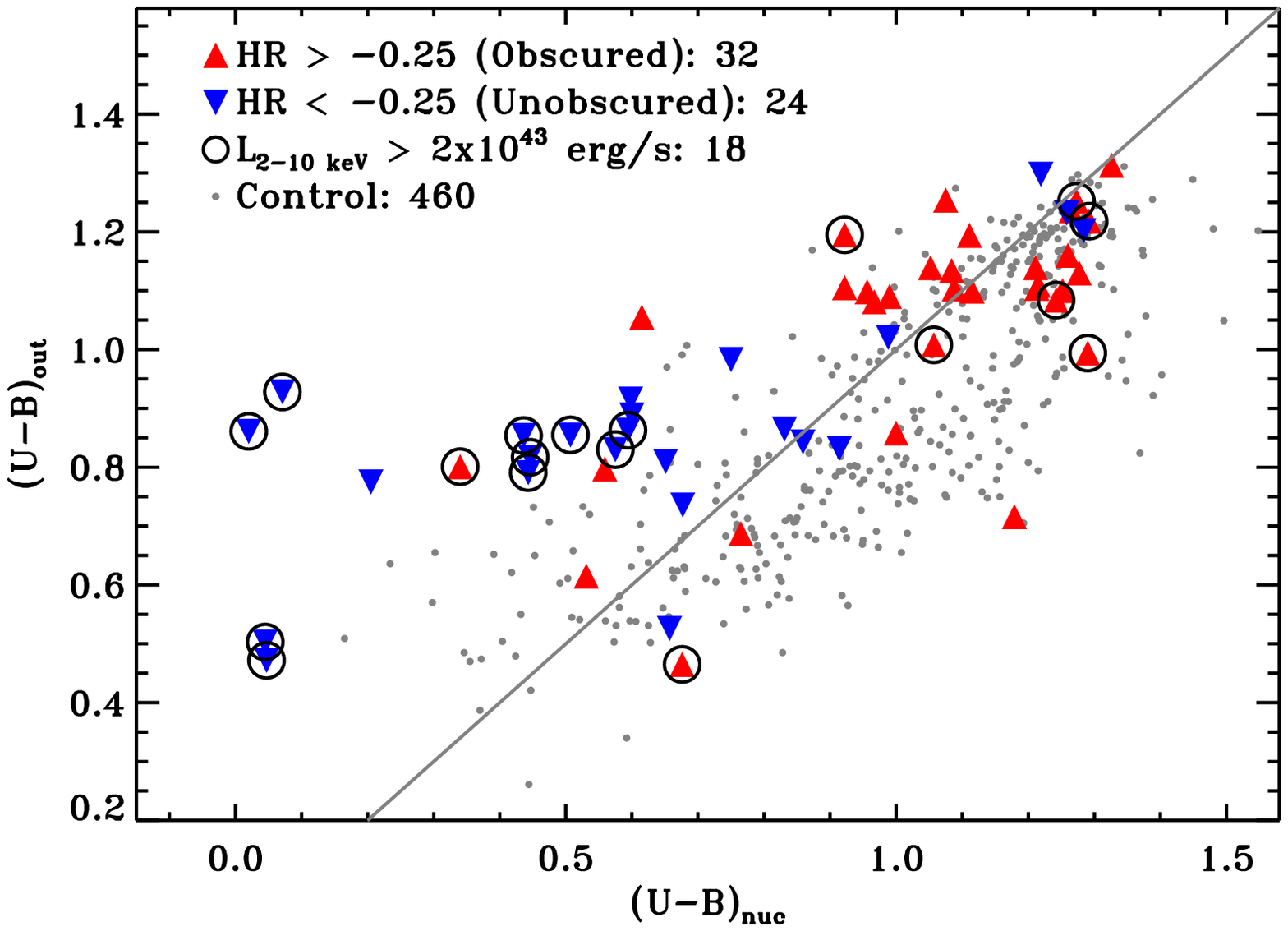}
\caption{Outer colour $(U-B)_{\rm out}$ vs.\ nuclear colour $(U-B)_{\rm nuc}$ for the AGN and control samples. Symbols are as in Fig.\ \ref{fig:ub_mstar}. The diagonal gray line shows where $(U-B)_{\rm out}=(U-B)_{\rm nuc}$. Most of the unobscured AGN host galaxies have relatively blue nuclear colours, while the obscured AGN host galaxies and the control galaxies exhibit roughly similar outer and nuclear colours.}
\label{fig:ub_out_ub_nuc}
\end{figure}

Whether considering the outer colours or the nuclear colours, we find that X-ray-unobscured AGNs are generally bluer than X-ray-obscured AGNs. We also find that the nuclear colours of galaxies hosting AGNs, particularly unobscured AGNs, are more often blue than the nuclear colours of the control sample. This is illustrated by Fig.\ \ref{fig:ub_out_ub_nuc}, which shows the outer colours vs.\ the nuclear colours of AGN hosts and control galaxies. Just like for the outer colours, the probability that the nuclear colours of the obscured and unobscured AGNs are drawn from the same parent population is small, as is the probability that the unobscured AGNs are drawn from the same parent population as the control galaxies. However, the nuclear colours of the obscured AGN hosts exhibit a 34 per cent probability of having been drawn from the same parent population as the control galaxies.

\begin{figure*}
\includegraphics[width=7in]{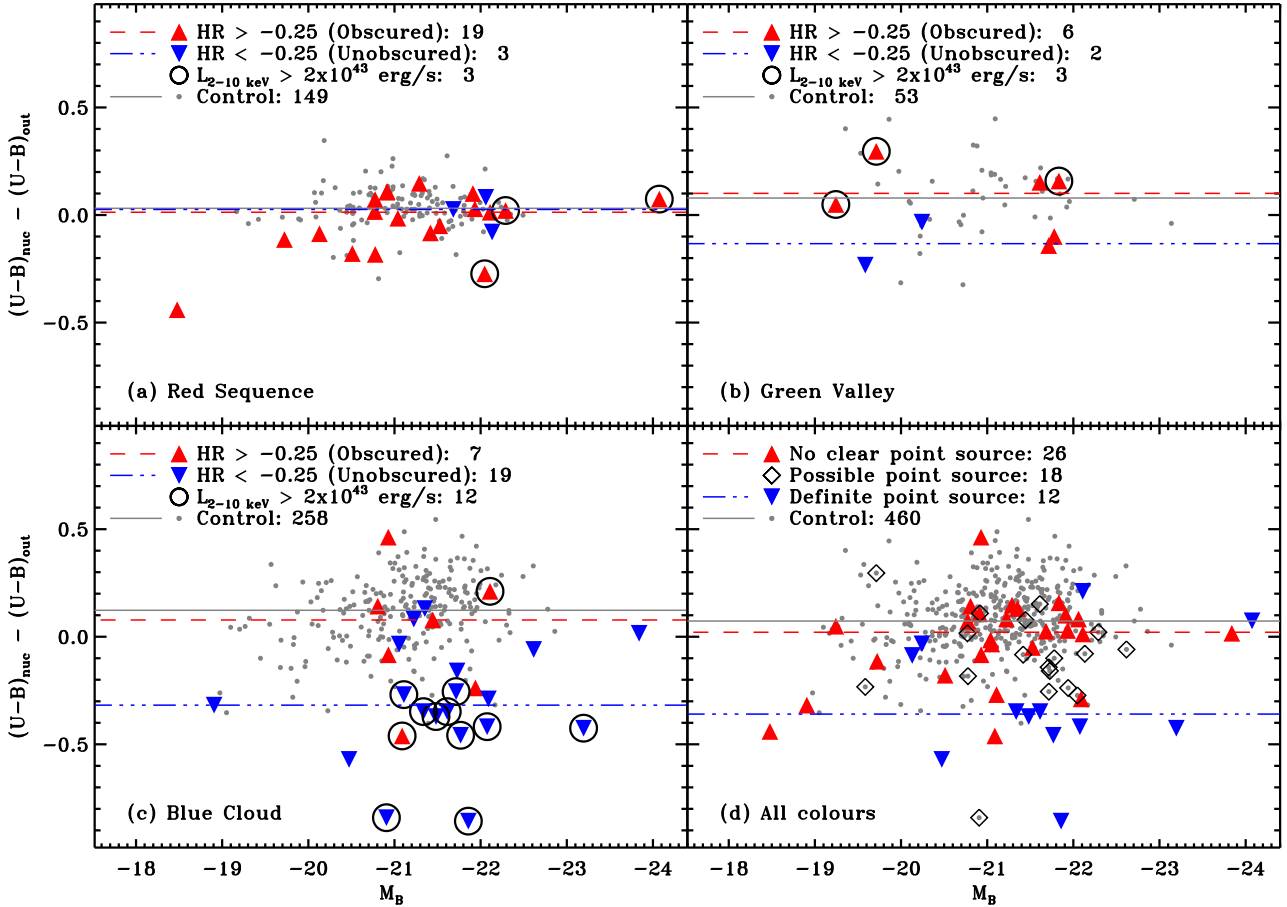}
\caption{Optical colour gradient vs.\ $B$ band magnitude. The number of objects in each sub-sample is indicated on the figure. {\it Panels (a)--(c)}: AGN and control sub-samples selected based on location in the optical CMD, as indicated on each panel; symbols are as described for Fig.\ \ref{fig:ub_mb}; encircled symbols indicate AGNs with X-ray luminosities $L_{\rm 2-10 \ keV} > 2 \times 10^{43}$ erg s$^{-1}$; lines represent the median colour gradients of the control sample (gray solid line), unobscured AGNs (blue dash-triple dotted line) and obscured AGNs (red dashed line). {\it Panel (d)}: Symbol shape and colour indicate whether or not the galaxy image contains a visibly bright point source (see Fig.\ \ref{fig:ps_visibility} for examples); lines represent the median colour gradients of the control sample (solid gray line), AGNs with a definite point source (blue dash-triple dotted line) and AGNs with no clear point source (red dashed line). Definite point sources [inverted blue triangles on panel (d)] are correlated with large negative colour gradients and unobscured AGNs [inverted blue triangles on panels (a)-(c)].}
\label{fig:delta_ub_mb_hr_ps}
\end{figure*}

We further find that the median colour gradients [defined here as $\Delta(U-B) = (U-B)_{\rm nuc} - (U-B)_{\rm out}$] differ between the control galaxies and the X-ray hosts, as shown in Fig.\ \ref{fig:delta_ub_mb_hr_ps}, where we separate the X-ray hosts and control galaxies into sub-samples according to optical colour and X-ray hardness ratio [panels (a)-(c)] or visibility of an optical point source [panel (d)]. Red-sequence AGN hosts exhibit an almost non-existent colour gradient for all levels of X-ray obscuration. Green-valley and blue-cloud galaxies hosting obscured AGNs exhibit nuclear colours that are typically redder than their outer regions. Galaxies of all colours that host obscured AGNs exhibit colour gradients that are very similar to the colour gradients shown by the control sample galaxies (small gray circles, solid line).

Panels (a)--(c) of Fig.\ \ref{fig:delta_ub_mb_hr_ps} demonstrate a clear contrast between the colour gradients of galaxies hosting obscured AGNs and those hosting unobscured AGNs. In general, the nuclei of galaxies hosting unobscured AGNs (blue inverted triangles) are estimated to be significantly bluer than their outer regions, and the difference increases as the integrated optical colours change from red to blue. The median colour gradients of the control sample galaxies shows that those in the blue cloud exhibit the strongest colour gradients, such that the nuclei are redder than the outer regions of the galaxies, which is the opposite of what we find among blue-cloud AGN hosts. A similar colour gradient is also evident among green-valley control sample galaxies, though to a lesser extent. Linear fits to the colour gradients of the control sample and the two AGN sub-samples appear to depend slightly on both redshift and $B$ band magnitude, but due to the scatter present in the sub-samples, we do not consider these to be significant results.

The disparity between the colour gradients of the unobscured AGN hosts and the control galaxies in the blue cloud are quantitatively confirmed by a two-dimensional K--S test, which indicates a probability $<<1$ that the two samples are drawn from the same parent population. The median control galaxy colour gradient at $\Delta(U-B) > 0$ further emphasizes the result that the outer regions of control sample galaxies are typically bluer than the nuclear regions. In contrast, the AGN sample exhibits a wider distribution of colour gradients between the nuclear and outer regions, extending to $\Delta(U-B) \sim -0.8$.

Higher X-ray luminosities ($L_{\rm 2-10 \ keV} > 2 \times 10^{43}$ erg s$^{-1}$), indicated by encircled symbols in panels (a)--(c) of Fig.\ \ref{fig:delta_ub_mb_hr_ps}, appear to be associated with stronger colour gradients for unobscured AGN hosts. The high-luminosity X-ray-{\it unobscured} AGNs exhibit gradients more negative than or consistent with the average for the unobscured sub-sample, indicating that the nuclei are particularly blue compared to the outer colours. In addition, half of the high-luminosity X-ray-obscured AGNs have gradients that are larger than the median for the obscured sub-sample, indicating that the nuclei are particularly red compared to the outer regions. As discussed in Section \ref{samples:agn}, the X-ray luminosities represent lower limits for the obscured AGNs, thus two of the `lower luminosity' X-ray-obscured AGNs that have colour gradients slightly {\it lower} than the average for that sub-sample [e.g.\ $\Delta(U-B) \sim -0.1$] have {\it intrinsic} X-ray luminosities estimated to be slightly in excess of $L_{\rm 2-10 \ keV} = 2 \times 10^{43}$ erg s$^{-1}$.

Finally, we find that the outer regions of galaxies hosting unobscured AGNs typically exhibit colours that are redder than the nuclear regions [$\Delta(U-B) < 0$] and have a very low probability of having been drawn from the same parent population as the control sample. Obscured AGNs exhibit nuclear and outer colours similar to those of the control galaxies, but with a wider distribution in $\Delta(U-B)$, demonstrating that even hosts of obscured AGNs exhibit different colour gradients than typical control galaxies. One final K--S test indicates that the colour gradients of the obscured and unobscured AGN samples (independent of the integrated colours of the host galaxies) have only a 0.4 per cent probability of having been drawn from the same parent population.

In panel (d) of Fig.\ \ref{fig:delta_ub_mb_hr_ps} we again present the colour gradients, now indicating the visibility of a point source, following the criteria described in Section \ref{data:optical}. Most of the clearly identified point sources (blue inverted triangles) are located in galaxies with particularly blue nuclear colours and are generally the same galaxies that are X-ray-unobscured and X-ray-luminous [compare to panels (a)--(c)]. AGNs for which we are unable to identify a point source generally coincide with X-ray-obscured AGN host galaxies. This suggests that very blue nuclear colours may reliably indicate the presence of visible nuclear point sources. The median colour gradients of the control sample and the AGN samples having either clearly identified point sources or no clear point source are relatively similar to the median gradients of the AGN and control galaxy samples shown in panels (a)--(c).

Furthermore, panel (d) of Fig.\ \ref{fig:delta_ub_mb_hr_ps} shows that if an X-ray-selected AGN has an obvious point source, it is likely to have very {\it different} nuclear and outer colours (large absolute gradients), and if an X-ray-selected AGN does {\it not} have an obvious point source, it is likely to have {\it similar} nuclear and outer colours. Panels (a)--(c) of Fig.\ \ref{fig:delta_ub_mb_hr_ps} indicate that the highest levels of potential AGN light contamination of observed optical colours seem to correlate with unobscured AGNs (HR $< -0.25$). Thus, AGN colour contamination seems most likely to occur in AGN host galaxies that have visible point sources {\it and} low X-ray obscuration. Finally, we find very few control galaxies that exhibit colour gradients similar to those of the luminous unobscured AGN host galaxies and are not already identified as AGNs.

Overall, Figs \ref{fig:hist_ub_out} -- \ref{fig:delta_ub_mb_hr_ps} indicate that galaxies hosting {\it unobscured} X-ray-luminous AGNs tend to have high current or recent star formation rates. Though the very blue nuclear colours (Fig.\ \ref{fig:hist_ub_nuc}) of these galaxies may be influenced by the AGNs, the significant colour gradients shown in Fig.\ \ref{fig:delta_ub_mb_hr_ps} suggest that the blue colours of the {\it outer} regions (Fig.\ \ref{fig:hist_ub_out}) accurately characterize the dominant stellar populations, while the nuclear regions are alone contaminated by AGN light. On the other hand, {\it obscured} X-ray-luminous AGNs tend to be hosted by intrinsically red galaxies with even redder nuclear regions, suggesting minimal current or recent star formation. In general, hosts of X-ray-obscured AGNs exhibit colour gradients similar to those of the control sample galaxies, with nuclear colours similar to or slightly redder than the outer regions.

The colours and colour gradients of the control galaxies and hosts of X-ray-unobscured AGNs may be explained in terms of the expected distribution of star formation in the galaxies. Younger (bluer) stars would be expected in the discs of non-AGN hosts more so than in the bulges, while the activity fueling significant black hole growth may at the same time contribute to star formation in the nuclear regions of an AGN host galaxy. We could then conclude that X-ray-obscured AGNs are undergoing minimal star formation either in the nuclear regions or in the outer regions.

%% Summary %%
Based on Figs \ref{fig:ub_mb}-\ref{fig:delta_ub_mb_hr_ps}, we find that the integrated and outer colours of the control galaxies possess a bimodality roughly consistent with the colour bimodality demonstrated by the general population of galaxies at $z \sim 1$ (e.g.\ Willmer et al.\ 2006; Faber et al.\ 2007). In contrast, the control galaxy {\it nuclear} colours feature a red peak with a blue tail. We further find a colour gradient with a single broad distribution peaking near $\Delta(U-B) \sim 0.07$, in agreement with Koo et al.\ (2005) who found that the bulges of late-type galaxies at redshifts $z \sim 1$ are typically redder than the accompanying discs.

We find that the integrated, outer, and nuclear colours of AGN host galaxies correlate with the level of X-ray obscuration experienced by the AGNs, in that blue-cloud AGN hosts typically suffer less obscuration than red-sequence AGN host galaxies. This is the opposite of what one might expect from consideration of the interstellar medium, which is thought to be more prevalent in blue-cloud galaxies than in red-sequence galaxies. However, the material obscuring the X-ray emissions from an AGN may be located in a more concentrated region near the black hole, which may then indicate a problem with models that predict that AGN feedback would remove such material from the nuclear regions before (or while) shutting down star formation in the outer regions of the galaxy. As mentioned above, we also find that the integrated colours of red-sequence AGN hosts are bluer than those of red-sequence control galaxies, and the integrated colours of blue-cloud AGN hosts are redder than those of blue-cloud control galaxies, possibly indicating a connection between the presence of an AGN and the star formation in the host galaxy (see, e.g., Schawinski et al.\ 2007; Georgakakis et al.\ 2008).

X-ray obscuration also correlates with AGN host galaxy colour gradients. Though red-sequence AGN host galaxies are typically obscured, both those that are obscured and those that are unobscured have colour gradients very similar to the red-sequence control galaxies ($\Delta U-B \sim 0$). Obscured AGNs in {\it blue-cloud} galaxies have colour gradients $\Delta U-B \sim 0.1$, which is again very similar to the blue-cloud control galaxies. In contrast, the X-ray-unobscured AGNs hosted by blue-cloud galaxies have nuclear regions that are noticeably bluer than the outer regions ($\Delta U-B \sim -0.25$). The colour gradient decreases (indicating even bluer nuclei) for high-luminosity AGNs, which is due in large part to the visibility of the AGN for many of these systems, as indicated in panel (d) of Fig.\ \ref{fig:delta_ub_mb_hr_ps}. Finally, we also observe that the luminous {\it obscured} AGNs have slightly redder colour gradients than the less luminous obscured AGNs.

%%%%%%%%%%%%%%%%%%%%%%%%%%%%%
\subsection{NUV$-R$ integrated colours}\label{results:nuvr}
As discussed in Section \ref{intro}, UV-optical colours are particularly good indicators of recent star formation, but they may also be rather sensitive to contamination from AGN light. In the preceding section, we presented the nuclear and outer colours of AGN host galaxies. We now compare these colours to the {\it integrated} UV-optical colours of the same galaxies to explore the effect of an AGN on the measured galaxy colours and the estimated star formation rates. In particular, we are concerned with the possibility that AGN host galaxies may appear too blue, which is particularly relevent for AGNs that are identifiable as visible point sources.

Work by Kauffmann et al.\ (2007), using a low-redshift sample of optically selected AGNs, established a lack of colour contamination among the AGN host galaxies in their sample (Section \ref{intro}). However, the AGNs considered in the current work are X-ray-selected and at higher redshifts ($z \sim 1$). The presence of visible point sources in many of our AGN host galaxies [cf.\ panel (d) of Fig.\ \ref{fig:delta_ub_mb_hr_ps}] underscores the importance of checking this particular sample for potential contamination of AGN host galaxy integrated colours.

Integrated UV-optical colours are unavailable for certain galaxies in the AGN and control samples. In addition, a few of the {\it available} UV-optical colours are considered unreliable due to high values of $\chi^{2}$ (cf.\ Section \ref{data:nuvr}). However, as discussed in Section \ref{samples:cmds}, the galaxies for which we {\it do} have reliable UV-optical colours are representative of the full sample.

\begin{figure}
\includegraphics[width=3.5in]{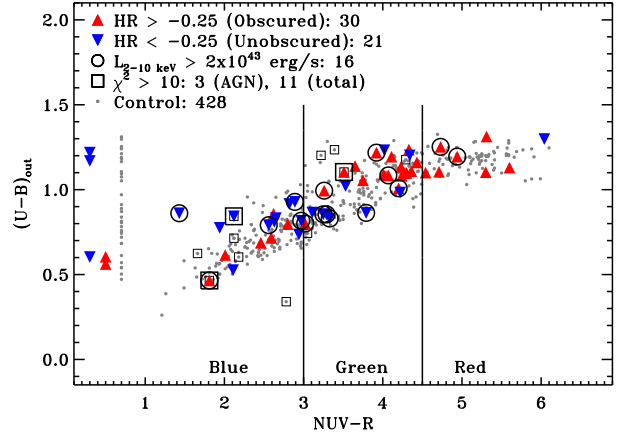}
\caption{Outer optical colours vs.\ integrated UV-optical colours. Symbols are as described for Fig.\ \ref{fig:delta_ub_mb_hr_ps}, except boxed symbols indicate galaxies for which the NUV$-R$ colours are not considered reliable. UV-optical colours are not available for galaxies represented by symbols at NUV$-R < 1$. Vertical lines separate the figure into NUV$-R$ red, green and blue colour regions. The number of objects in each sub-sample is indicated on the figure. The outer optical colours and integrated UV-optical colours exhibit a tight correlation, indicating that an AGN typically does not contaminate the measured integrated colours of its host galaxy.}
\label{fig:ub_out_nuvr_hr}
\end{figure}

\begin{figure}
\includegraphics[width=3.5in]{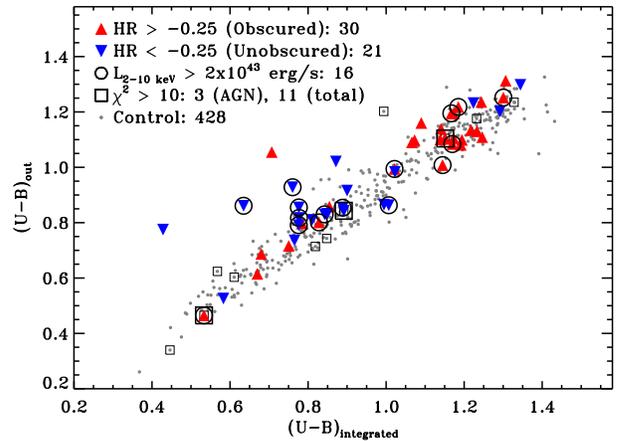}
\caption{Outer optical colours vs.\ integrated optical colours. Samples and symbols are as described for Fig.\ \ref{fig:ub_out_nuvr_hr}. As in Fig.\ \ref{fig:ub_out_nuvr_hr}, the close correlation between the outer and integrated colours indicates a general lack of contamination from AGN light.}
\label{fig:ub_out_ub_hr}
\end{figure}

Fig.\ \ref{fig:ub_out_nuvr_hr} presents the outer optical colours $(U-B)_{\rm ext}$ and integrated UV-optical colours NUV$-R$. We observe a tight correlation between the outer optical colours, which are expected to accurately represent the intrinsic colours of AGN host galaxies, and the integrated UV-optical colours. From this we conclude that, in general, an active nucleus does {\it not} significantly affect the integrated colours of an AGN host galaxy, even when near-UV colours are used. The results are similar for integrated optical colours, as we show in Fig.\ \ref{fig:ub_out_ub_hr}, though five AGNs scatter off the ridgeline toward bluer integrated colours (OBJNOs 12012471, 13026061, 13042389, 12028367, and 12012474), three of which (OBJNOs 12012472, 12028367, and 12012474) show probable nuclear point sources.

There are three AGN host galaxy outliers in Fig.\ \ref{fig:ub_out_nuvr_hr} (OBJNOs 13026061, 12012471, and 13035123), one of which has unreliable UV-optical colours (OBJNO 13035123); compared to their outer optical colours, they all seem to be slightly too blue in the UV-optical colours. The most X-ray luminous of the three (OBJNO 13026061) is also the most deviant from the colour-colour correlation. The single {\it control} galaxy that deviates considerably from the correlation (OBJNO 12008055) has an unreliable UV-optical colour. Based on its optical spectrum, it is classified as a Seyfert galaxy, but it is outside of the {\it Chandra} X-ray survey region, so its X-ray luminosity is unknown.

We observe a near separation between the UV-optical colours of obscured and unobscured AGNs at NUV$-R = 3.5$ in Fig.\ \ref{fig:ub_out_nuvr_hr}; obscured AGN host galaxies (red triangles) exhibit redder colours, and unobscured AGN host galaxies (blue triangles) exhibit bluer colours. This separation is especially true for the more X-ray luminous AGNs, while the colours of some of the lower luminosity AGNs are unexpected with respect to the level of X-ray obscuration. In particular, one of the reddest hosts has a very low level of obscuration (OBJNO 12020028; HR $ = -0.49$). In general, the blue obscured AGNs are not strongly obscured (HR $ < 0.03$), with one exception (OBJNO 12016316; HR $ = 0.77$). Thus, for both the aperture colours and the integrated colours we observe the trend that {\it unobscured AGNs tend to be hosted by blue galaxies and obscured AGNs tend to be hosted by red galaxies.}

% --> The following fractions are restricted to objects for which chi2 < 10:
% --> Red: 8/92 = 8.7%
% --> Green: 25/170 = 14.7%
% --> Blue: 15/153 = 9.8%
% --> Total: 48/415 = 11.6%
Since we use a control sample designed to match the intrinsic characteristics of the AGN host galaxies (Section \ref{samples:control}), we can estimate the fraction of UV-optical red-sequence, green-valley and blue-cloud galaxies that are identified as AGNs. We identify $8.7^{+4}_{-3}$ per cent of the red-sequence galaxies as X-ray-selected AGNs; the fraction of blue-cloud galaxies that are selected as AGNs comes in next at $9.8\pm{3}$ per cent; AGNs are most prevalent in the green valley where $14.7^{+4}_{-3}$ per cent of the control sample galaxies host X-ray-selected AGNs. The differences between these fractions are only suggestive as they are all consistent with one another, so we conclude that AGN are present in roughly 12 per cent of the control galaxies, regardless of the integrated UV-optical colour.

\begin{figure}
\includegraphics[width=3.5in]{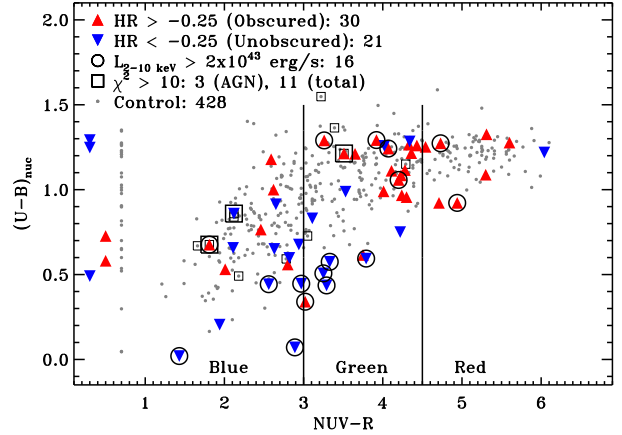}
\caption{Nuclear optical colours vs.\ integrated UV-optical colours. Samples, symbols and lines are as described for Fig.\ \ref{fig:ub_out_nuvr_hr}. In contrast to the results shown by Figs \ref{fig:ub_out_nuvr_hr} and \ref{fig:ub_out_ub_hr}, we find here a very loose correlation between the integrated and nuclear colours, indicating a potentially large contamination by the AGN light.}
\label{fig:ub_nuc_nuvr_hr}
\end{figure}

Finally, in Fig.\ \ref{fig:ub_nuc_nuvr_hr} we present the relationship between the {\it nuclear} $U-B$ colours and the integrated UV-optical colours. The control galaxies (small gray circles) exhibit a loose correlation between the nuclear and integrated colours, such that the nuclear colours redden with the integrated colours until NUV$-R \sim 4.5$, where the optical nuclear colours saturate as the UV-optical colours continue to redden. The host galaxies of the X-ray-obscured and lower luminosity X-ray-unobscured AGNs generally exhibit the same loose correlation and have colours consistent with the control sample galaxies. However, the higher luminosity, X-ray-unobscured AGNs ($L_{\rm 2-10 \ keV} > 2 \times 10^{43}$ erg s$^{-1}$; encircled inverted triangles) exhibit very blue nuclear colours, supporting our earlier conclusion that these AGNs show noticeable blue point sources, which do not significantly affect the integrated colours.

In conclusion, X-ray obscuration and/or luminosity appear to correlate with the nuclear colours of our AGN host galaxies. Higher luminosity, unobscured AGNs have blue nuclear colours, as mentioned above. The lower luminosity, X-ray-unobscured AGNs generally have redder nuclear colours than the higher luminosity systems, but bluer nuclear colours than galaxies hosting obscured AGNs. Among the X-ray-obscured AGNs, there is not a clear correlation between nuclear colour and X-ray luminosity.

%%%%%%%%%%%%%%%%%%%%%%%%%%%%%
% Discussion
%%%%%%%%%%%%%%%%%%%%%%%%%%%%%
\section{Discussion}\label{discussion}

%%%%%%%%%%%%%%%%%%%%%%%%%%%%%
\subsection{AGN host galaxy colours}\label{disc:colors}
Several of the figures presented in the previous section suggest a correlation between host galaxy colours (optical {\it and} UV-optical) and X-ray obscuration. In particular, we find that red galaxies typically host obscured AGNs while blue galaxies host unobscured AGNs, suggesting a connection between the galaxy's star formation history and the cloud obscuring the central accretion disc. If the outer regions of a galaxy are still dominated by young stars when the AGN is unobscured, then the energetic feedback often predicted to clear the nuclear regions and halt star formation is clearly not functioning as expected. The nuclear regions have been cleared, but star formation continues. Thus, we do not yet sufficiently understand the timescales during which these processes operate.

As an alternative to a correlation between significant AGN obscuration and rapidly aging (and hence reddening) stellar populations, some of our galaxies may be experiencing dust-enshrouded star formation, with the host galaxies playing a role in obscuration of the soft X-rays emanating from the nuclear regions (e.g.\ Rigby et al.\ 2006). In this case, we might expect similar star formation rates in both the red and blue AGN host galaxies. However, many authors (e.g.\ Bell et al.\ 2004; Silverman et al.\ 2005; Georgakakis, Georgantopoulos \& Akylas 2006; Rovilos \& Georgantopoulos 2007) have found that high fractions of optically red galaxies, even those hosting luminous X-ray sources, exhibit early-type morphologies, in which we do not expect to find significant dust-enshrouded star formation. Visual inspection of the {\it HST}/ACS images of our red-sequence AGN host galaxies confirms that most do indeed exhibit early-type morphologies, although one (OBJNO 12004450) also features a prominent dust lane and another appears to have a fairly prominent blue disc (OBJNO 12025302) around a red nucleus.

%%%%%%%%%%%%%%%%%%%%%%%%%%%%%
\subsection{Previous observational analyses}\label{disc:observations}

Kauffmann et al.\ (2007) measured the host galaxy optical and UV-optical colours of a low-redshift ($z \sim 0.05$) AGN sample selected using optical spectroscopy. Taking advantage of the ability to measure the nuclear spectra of a nearby sample, they estimated the ages of the nuclear stellar populations [using $D_{n}$(4000) and H$\delta_{A}$] and the AGN accretion rates (using L[OIII]). In addition, they measured the integrated optical ($g-r$) and UV-optical (NUV$-r$) galaxy colours and considered the optical ($g-i$) colour profiles. Galaxies hosting strongly accreting AGNs contained young stellar populations in the outer regions (as determined by blue $g-i$ colours) {\it and} in the nuclear regions (as determined by H$\delta_{A}$); Kauffmann et al.\ (2007) found a strong correlation between the accretion rate and the age of the {\it nuclear} stellar population. However, they also determined that the nuclear {\it colours} were independent of the AGN accretion rates, and that the integrated UV-optical colours were not correlated with the age of the nuclear stellar population. Furthermore, they found that a young stellar population in the outer region does not correlate with either a young nuclear stellar population or a high AGN accretion rate.

Even though the sample discussed here (at $z \sim 1$ and X-ray-selected) differs from the sample used by Kauffmann et al.\ (2007), the two sets of results are consistent in that we also find that the integrated UV-optical colours are only loosely correlated with the nuclear optical colours of the galaxies. In the context of the present study, we take this to mean that the integrated UV-optical colours are not biased by the nuclear regions (particularly the accreting black hole) of most AGN host galaxies. Thus, what Kauffmann et al.\ (2007) showed for the hosts of low-redshift, optical spectroscopically selected AGNs, we have shown to also be true for their higher-redshift, X-ray-selected counterparts.

%%%%%%%%%%%%%%%%%%%%%%%%%%%%%
\subsection{Comparisons to models}\label{disc:models}
We now compare our results to the three models of AGN feedback described in Section \ref{intro} and the effects on star formation in the host galaxy. For each of the models, we summarize the proposed scenario and then compare it to the colour gradients presented in Section \ref{results:ub}, along with the AGN X-ray luminosities and obscuration.

\subsubsection{Mergers and quasar feedback}
In a pair of papers, Hopkins et al.\ (2008a, 2008b) described the results of applying the following two hypotheses to cosmological simulations: (1) major mergers between gas-rich galaxies trigger quasars, though they may also be triggered by alternate processes (Hopkins et al.\ 2008a), and (2) such mergers will lead to short-term quenching of star formation (Hopkins et al.\ 2008b). The expected applicability of the models presented by Hopkins et al.\ (2008a, 2008b) changes with redshift. AGNs (specifically, quasars) at redshifts $z\ ^{>}_{\sim}\ 1$ are typically caused by mergers, while secular mechanisms, such as disc instabilities, activate black hole growth at lower redshifts ($0 < z < 0.5$).

Results based on the first hypothesis suggest a rapid progession from an optically blue, star-bursting galaxy hosting an obscured quasar to a quenched red-sequence galaxy briefly hosting an unobscured, dying quasar (Hopkins et al.\ 2008a). Consideration of the second hypothesis led the authors to conclude that while the starburst accompanying a merger of gas-rich galaxies will quickly consume enough gas to quench itself (leaving a red merger remnant), the result will be temporary, and some form of feedback will be necessary to maintain the colour. Feedback from, for example, a quasar (as described by Hopkins et al.\ 2005a, 2005b), shocks, tidal heating or winds driven by starbursts could provide the energy necessary to maintain a high gas temperture for an outer period of time, thereby at least partially accounting for the observed red-sequence population (Hopkins et al.\ 2008b).

This scenario seems to explain at least one of our AGN host galaxies quite well. The $z \sim 1$ galaxy (OBJNO 12020452) exhibits blue outer and integrated colours [$(U-B)_{\rm out} = 0.48$ and $(U-B)_{\rm integrated} = 0.53$ in Fig.\ \ref{fig:ub_out_ub_hr}] and has a high X-ray luminosity ($L_{\rm 2-10 \ keV} = 1.8 \times 10^{44}$ erg s$^{-1}$), yet the AGN itself is {\it obscured} (HR$ = 0.01$). The {\it HST}/ACS images show a prominent offset nucleus with a large, possibly dusty, region extending to one side, suggestive of an ongoing interaction or merger. The mild X-ray obscuration and young stellar population suggest that the galaxy has not (yet) experienced any significant feedback from the AGN, putting it in an early stage of the scenario presented by Hopkins et al.\ (2005a, 2005b, 2008a, 2008b).

Although Fig.\ \ref{fig:ub_mb} is consistent with the result presented by Hopkins et al.\ (2008a) that AGN hosts tend to be red (or located towards the red edge of the blue cloud), this figure also demonstrates a potential challenge to the final stages of the merger scenario, as it is described by Hopkins et al.\ (2005a, 2005b, 2008a). In our sample, most of the AGN host galaxies located on the red sequence are X-ray obscured, which is inconsistent with a prediction that feedback from the AGN forced the gas supply away from the nuclear regions while (approximately) simultaneously quenching star formation. However, Bundy et al.\ (2008) observed that local, optical spectroscopically selected AGNs may exhibit greater consistency with this scenario than higher-redshift, X-ray-selected AGNs.

\subsubsection{Maintenance of giant elliptical galaxies}
Specifically addressing a cycle that giant elliptical galaxies may undergo, Ciotti \& Ostriker (2007) presented a scenario that intimately links host galaxy stellar populations with black hole growth and AGN feedback. In their picture, supernovae release metal-rich gas from aging stellar populations, about half of which is caught up in galactic winds, while a radiative instability causes about half to collapse toward the nuclear regions and feed a central starburst; the remaining small fraction of the gas (estimated at $<1$ per cent) feeds the central SMBH. Though initially obscured (Compton-thick; $N_{H} > 10^{24}$ cm$^{-2}$), the AGN and nuclear starburst are gradually revealed as they consume the gas. Finally, an expanding central hot bubble, created by energetic feedback from the AGN, is expected to shut down the black hole growth and star formation. A luminous AGN is expected to be briefly visible before fading to a low-luminosity AGN in a galaxy with a mixed-age stellar population. This cycle may repeat after the nuclear regions have cooled sufficiently. 

With a predicted duty cycle of only a few per cent, we expect to detect no more than one X-ray-luminous, unobscured AGN (i.e., a system in the predicted QSO phase) in our sample. Though we observe one red-sequence AGN host that has an apparent point source (OBJNO 13009690), it is X-ray-obscured and thus does not quite fit the predictions of this scenario. However our lack of observing systems during the QSO phase predicted by this model is not surprising, given the predicted rarity of such systems, and probably cannot be used to rule out the effectiveness of this scenario at explaining the relationships between AGN and their host galaxies.

We do not expect to detect Compton-thick AGNs with the AEGIS {\it Chandra} survey, so such objects would not be included in our AGN sample. However, they may be included in the control sample, because they would possess redshifts and stellar masses similar to those of AGN host galaxies. AGNs that are obscured but not quite Compton-thick (e.g.\ $10^{22}$ cm$^{-2} < N_{H} < 10^{24}$ cm$^{-2}$) should contribute to our sample of X-ray-selected AGNs, exhibiting low X-ray luminosities, significant obscuration of soft X-rays and red nuclear regions. As shown by Figs \ref{fig:hist_ub_nuc} and \ref{fig:delta_ub_mb_hr_ps}, our sample contains several systems that match this description.

Continuing to follow the scenario described by Ciotti \& Ostriker (2007), systems that we would observe as QSOs are expected to exhibit high X-ray luminosities, low X-ray obscuration and blue nuclear colours. As indicated by the encircled, blue, inverted triangles in panel (c) of Fig.\ \ref{fig:delta_ub_mb_hr_ps}, our AGN sample certainly contains systems fitting this description. However, for many of these systems it is not certain that the blue nuclear colours result entirely from the presence of young nuclear stars. We expect the AGN light to bias the nuclear colours of the most X-ray-luminous, unobscured AGNs, but several of the low-luminosity, unobscured AGNs also exhibit blue nuclear colours. This may represent a stage part way between the initial QSO and the final E$+$A stage, in which star formation has stopped, but the stellar populations are still young. Unobscured AGNs exhibiting red nuclear colours may correspond to the final population of galaxies exhibiting low X-ray luminosities and E$+$A spectra.

We may thus interpret Fig.\ \ref{fig:delta_ub_mb_hr_ps} in the following manner:
\begin{enumerate}
\item The lower X-ray luminosity, obscured systems (uncircled red triangles) represent recently triggered AGNs. They are heavily obscured, but not Compton-thick. Central starbursts subjected to heavy obscuration lead to the typically red nuclear colours.
\item Obscured but highly X-ray luminous AGNs (encircled red triangles) may represent a slightly later stage during which the AGNs are approaching peak growth, but have not yet caused sufficient feedback to shut down star formation and significant black hole accretion.
\item Unobscured AGNs exhibiting high X-ray luminosities (encircled blue inverted triangles) have recently caused the expansion of a central hot bubble, shutting down the star formation and AGN growth, but briefly revealing the AGN as a QSO. The nuclear light of the host galaxies may suffer contamination from the AGN light, making it difficult to determine the definite ages of the nuclear stellar population.
\item Low X-ray luminosities in unobscured AGNs (uncircled blue inverted triangles) would indicate the final stage of the AGN cycle, in which the black hole growth has significantly slowed and the stellar population has begun to age and redden, explaining the redder nuclear colours of some of the host galaxies. Three of the red-sequence AGN hosts represented in Fig.\ \ref{fig:ub_mb} (OBJNOs 13025417, 13027442, and 12020028, in order of decreasing $B$ magnitude) exhibit low X-ray luminosities and obscuration, making them candidates for this stage of the scenario.
\end{enumerate}

The model presented by Ciotti \& Ostriker (2007) seems consistent with the characteristics of a sub-sample of the observed AGN host galaxies, but does not fully explain either every characteristic or every AGN host. For example, several of the unobscured AGN host galaxies represented in Figs \ref{fig:hist_ub_out} and \ref{fig:delta_ub_mb_hr_ps} exhibit blue outer colours. This scenario does not provide an obvious explantion for the apparent star formation in the outer regions of the galaxy, particularly if the AGN is triggered by gas collapse toward the {\it nuclear} regions. We may expect that expansion of the central hot bubble connects the X-ray obscuration of the AGN with the nuclear colours, but it would not be expected to significantly affect the outer colours.

\subsubsection{`Radio mode' accretion and feedback}
Croton et al.\ (2006) presented two complementary scenarios for AGN accretion and feedback, one of which (the `quasar mode') is very similar to the model discussed by Hopkins et al.\ (2005a, 2005b, 2008a), so we will focus our comparison to their work on `radio mode' accretion and feedback. This scenario refers to the central black hole of a static hot gas halo accreting material at a rate significantly less than the Eddington limit. It is most important to black hole growth at redshifts $z < 2$, when quasar mode accretion is believed to decrease in importance (see fig.\ 3 of Croton et al.\ 2006). Based on their findings, Croton et al.\ suggested that feedback from radio mode accretion may be enough to halt star formation in the host galaxy, allowing the stellar populations to age and redden. Though not explicitly predicted from the radio mode models, we would expect such systems to exhibit red optical colours (nuclear and outer) and relatively low X-ray luminosities, due to the low rate of accretion of material on to the black hole.

Figs \ref{fig:ub_out_nuvr_hr} and \ref{fig:ub_nuc_nuvr_hr} show that our sample contains AGN host galaxies that appear to be consistent with the scenario described by Croton et al.\ (2006). Several obscured AGN host galaxies exhibit low X-ray luminosities ($L_{\rm 2-10 \ keV} < 10^{43}$ erg s$^{-1}$) and red optical colours in both the nuclear and outer regions; most of these galaxies exhibit green-valley and red-sequence UV-optical colours. Clearly, this scenario is not meant to describe the observations of all AGN host galaxies at all times, but it does offer an additional explanation to supplement the models presented by Ciotti \& Ostriker (2007), Hopkins et al.\ (2008a, 2008b) and many others. However, current semi-analytic models, such as that described by Croton et al.\ (2006), do not provide enough details to allow for a conclusive comparison with the work presented here.

The models described above apply to different galaxy populations, such as isolated giant ellipticals and interacting gas-rich discs. Therefore, it is not surprising that none of the models accurately describe all the characteristics of observed AGNs. Instead, each model describes a sub-sample of the observed population. Because of the incredible variety of AGN host galaxies, it seems reasonable to expect that multiple explanations are needed to properly describe them all. In addition, the time scales of the processes involved seems quite important for comparisons between observations and the models.

%%%%%%%%%%%%%%%%%%%%%%%%%%%%%
\subsection{Incompleteness of the AGN sample}\label{disc:incompleteness}
AGNs can be selected by a variety of methods, such as the X-ray luminosity requirement used for the current study, characteristic emission lines or emission line ratios from optical spectra (e.g.\ Baldwin, Phillips \& Terlevich 1981), infrared colours (e.g.\ Lacy et al.\ 2004; Stern et al.\ 2005; Hatziminaoglou et al.\ 2005), radio powers (e.g.\ Condon 1992) and radio/mid-IR ratios (e.g.\ Drake et al.\ 2003; Donley et al.\ 2005). However, these methods do not provide identical AGN samples, complicating attempts to characterize a `typical' AGN host galaxy. Perhaps the most commonly identified problem with X-ray selection is that the accretion discs of many AGNs experience significant attenuation from an obscuring cloud, that is, they are Compton-thick ($N_{H} > 10^{24}$ cm$^{-2}$); the X-rays are absorbed by the intervening material and re-emitted at IR wavelengths. On the other hand, lower levels of obscuration may prevent spectral identification of AGNs without hindering X-ray detection (e.g.\ Rigby et al.\ 2006).

Donley et al.\ (2005) studied a sample of AGNs selected by the radio/mid-IR ratio (radio-excess), and they found that only 40 per cent of such AGNs are also detected by X-ray observations in the {\it Chandra} Deep Field--North ($> 1$ Ms). They found few differences between characteristics such as X-ray to optical ratios and column densities of X-ray-detected and weakly X-ray-detected samples, suggesting that the two samples represent intrinsically similar AGN populations. The `weakly X-ray-detected' sample was comprised of X-ray sources detected at a lower significance than was required for the initial sample; these sources were only detected after the positions of their host galaxies had been determined. The inability to detect 60 per cent of the radio-excess AGNs could be explained by the orientation of the accretion disc relative to an observer's line-of-sight to the galaxy, instead of intrinsic differences in the AGNs.

Though Donley et al.\ (2005) did not address the colours of the AGN host galaxies, if the detected and undetected AGN populations {\it are} intrinsically similar, then it would be reasonable to expect that the colour gradients of the undetected AGN hosts would be similar to the colour gradients of the AGN hosts discussed in the current study. This conclusion is further supported by the similar distributions of column densities shown by Donley et al.\ (2005), which are related to the X-ray hardness ratios presented in the current study. Thus the relationships between colour gradients and column densities should be similar to the relationships discussed in Section \ref{results} between colour gradients and hardness ratios.

By stacking X-ray observations of galaxies that are not individually detected by the 200 ks AEGIS {\it Chandra} survey, Georgakakis et al.\ (2008b) found significant samples of Compton-thick AGNs and partly obscured low-luminosity AGNs. The host galaxies of these X-ray-undetected AGNs exhibit colours that place them on the red sequence or in the region between the red sequence and the blue cloud, which is consistent with what is found for the AGNs that {\it are} detected by the X-ray surveys. This again suggests that despite the incompleteness of the X-ray selection technique due to X-ray obscuration, inclusion of the missing AGNs would probably not significantly affect the results presented here.

The AGN sample described in Section \ref{data} does not contain any spectroscopically identified QSOs (i.e.\ broad-line AGNs), though it does include eight AGNs that have $L_{\rm 2-10 \ keV} > 10^{44}$ erg s$^{-1}$, which has also been used to define QSOs (e.g.\ Barger et al.\ 2007). The lack of spectroscopically identified QSOs is due in part to the rarity of such objects at redshifts $z \sim 1$ and in part to our requirement for reliable spectroscopic redshifts. The DEEP2 Redshift Survey selects against blue, point-like sources (see e.g.\ Faber et al., in preparation, and Willmer et al.\ 2006), such as galaxies dominated by AGN light. If we included these sources in the current study, they would probably be located in the lower right corners of Figs \ref{fig:ub_mb} and \ref{fig:nuvr_mr}. We expect that the nuclear colours would appear blue, because the galaxies would be dominated by the AGN light, and that the hardness ratios would indicate very low levels of obscuration. If QSOs are simply more luminous versions of the AGNs currently identified as unobscured and X-ray-luminous, then although the host galaxies would exhibit blue outer colours ($(U-B)_{\rm out}<1$; cf.\ encircled, blue inverted triangles in Fig.\ \ref{fig:ub_out_nuvr_hr}), the nuclear colours would be even bluer ($\Delta(U-B)<0$; cf.\ Fig.\ \ref{fig:delta_ub_mb_hr_ps}), indicative of ongoing star formation in the nuclear regions of the galaxy. However, if QSOs comprise a unique population, with respect to the colour gradients, and the outer colours were accurately determined to be blue, then we could conclude that the QSOs represented a stage in which stars were still actively forming in the outer regions of the galaxy.

One final note regarding the completeness of our AGN sample. Five AGN host galaxies are excluded from part of the analysis because their spectra are such that we are unable to determine UV-optical colours (Sections \ref{data:nuvr}, \ref{results:nuvr}). However, the characteristics of these AGNs and their host galaxies do not follow an identifiable trend (particularly with respect to X-ray luminosity or obscuration) that would set them apart as a unique population of AGNs. Thus, although we are unable to analyse the {\it UV-optical} colours of these AGNs, from our analysis of the {\it optical} colours and other characteristics of the systems (e.g.\ X-ray luminosities and hardness ratios), we do not expect that the inclusion of them would significantly affect the results presented here.

%%%%%%%%%%%%%%%%%%%%%%%%%%%%%
\subsection{X-ray sources in red and blue galaxies}\label{disc:redblue}
Bundy et al.\ (2008) investigated the quenching of star formation in AGN host galaxies by studying the host galaxy colours, accretion rates and stellar masses of X-ray-selected AGNs. Blue galaxies host the majority of their most X-ray-luminous AGNs ($L_{\rm 2-10 \ keV} > 10^{44}$ erg s$^{-1}$), though they note that the AGNs may bias the measured host galaxy colours. Figs \ref{fig:ub_out_nuvr_hr}-\ref{fig:ub_nuc_nuvr_hr} support the result from Bundy et al.\ (2008) that very luminous X-ray-selected AGNs (here, $L_{\rm 2-10 \ keV} > 2 \times 10^{43}$ erg s$^{-1}$; encircled symbols) tend to have blue host galaxies. Supporting the caution from Bundy et al.\ (2008) about a possible colour bias, we also find that these blue AGN host galaxies exhibit significant colour gradients (Fig.\ \ref{fig:delta_ub_mb_hr_ps}), generally experience minimal soft X-ray obscuration and exhibit nuclear colours that are bluer than their integrated colours (Fig.\ \ref{fig:ub_nuc_nuvr_hr}).

The general colour distribution of AGN host galaxies presented here is also consistent with the results from Bundy et al.\ (2008), in that AGNs at all X-ray luminosities $L_{\rm 2-10 \ keV} > 10^{42}$ erg s$^{-1}$ are not preferentially distributed among either red or blue host galaxies. This finding, coupled with accretion rate estimates, led Bundy et al.\ (2008) to the conclusion that AGNs hosted by massive, red galaxies accrete at about the same range of rates as AGNs hosted by blue galaxies, suggesting that AGN feedback does not {\it cause} star formation quenching, though the two processes may be related. As an alternative explanation, it is possible that winds associated with star formation may remove some of the obscuring material in blue host galaxies, while the relative lack of star formation winds may allow AGNs in red galaxies to remain obscured.

%%%%%%%%%%%%%%%%%%%%%%%%%%%%%
% Summary
%%%%%%%%%%%%%%%%%%%%%%%%%%%%%
\section{Summary}\label{summary}

The primary objective for the current work is to investigate the effect of AGN light on measurements of the host galaxy's colours. Using aperture photometry applied to {\it HST}/ACS $V$ and $I$ band images, we measure nuclear and outer colours of X-ray-selected AGN host galaxies and a carefully selected sample of control galaxies. From the observed $V-I$ aperture colours, we determine rest-frame $U-B$ aperture colours and compare them to the rest-frame integrated NUV$-R$ colours of the same systems. In addition, we analyse sub-samples of the AGN host galaxies based on X-ray luminosity, X-ray obscuration and visibility of a point source. We find several important results.

\begin{enumerate}

\item The nuclear colours of AGN host galaxies are generally bluer than their outer colours. This is unlike the typical nuclear colours of our control sample galaxies, which tend to be redder than their outer colours. The result for the AGN sample is especially pronounced for a sub-sample of X-ray luminous, unobscured AGNs (cf.\ Fig.\ \ref{fig:delta_ub_mb_hr_ps}).

\item AGNs in the {\it bluest} host galaxies exhibit a range of X-ray obscuration and luminosity. In contrast, AGNs in the {\it reddest} host galaxies are likely to be X-ray luminous, and they are typically obscured (cf.\ Figs \ref{fig:hist_ub_out} and \ref{fig:delta_ub_mb_hr_ps}). Coil et al.\ (2009; see their fig.\ 6) also demonstrated this correlation between optical colour and X-ray obscuration.

\item The {\it outer} optical colours and integrated UV-optical colours of AGN host galaxies are generally consistent with the colours of the control sample galaxies (cf.\ Fig.\ \ref{fig:ub_out_nuvr_hr}). However, the {\it nuclear} optical colours of unobscured AGNs are generally much bluer than those of the control sample galaxies (cf.\ Fig.\ \ref{fig:ub_nuc_nuvr_hr}). This indicates that the integrated colours typically represent the outer colours, not the nuclear colours. We therefore conclude that integrated NUV$-R$ colours (as well as redder optical colours) are generally reliable, even for X-ray-luminous AGNs, unless the AGNs are luminous, unobscured, {\it and} exhibit large colour gradients.

\item Large colour gradients within AGN host galaxies correlate with visible point sources, and this is more pronounced for X-ray unobscured AGN [cf.\ panel (d) of Fig.\ \ref{fig:delta_ub_mb_hr_ps}]. The correlation between point source visibility and X-ray obscuration suggests that X-ray obscuration and optical obscuration either come from the same region or come from different regions with highly correlated obscuration properties.

\item `Red' AGN host galaxies (as determined by the integrated colours) are typically bluer than `red' control galaxies (cf.\ Fig.\ \ref{fig:hist_ub_out}). As these colours are not likely to be contaminated by blue light from the AGNs, the AGN host galaxies have stellar populations that are younger than typical red-sequence galaxies.

\item Models predicting that major mergers cause AGN activity may explain a small fraction of the AGNs discussed in the current work. However, it appears that the majority of observed AGNs require alternate explanations, such as gas collapse caused by radiative instabilities.

\end{enumerate}

This work has shown that even for high-luminosity AGNs at $z \sim 1$, the optical and UV-optical colours are not likely to be strongly biased, unless the AGN is very luminous, unobscured and/or visible as a point source in an optical image. We have further shown that X-ray obscuration of the accretion disc correlates with galaxy colour. By considering a combination of predictions from various models, we seem to be able to explain the majority of the AGN hosts discussed here, but more work is needed to fully understand the connections between significant nuclear activity and a galaxy's star formation history.

%%%%%%%%%%%%%%%%%%%%%%%%%%%%%
\section*{Acknowledgments}
CMP and JRP acknowledge support from NASA ATP grant NNX07AG94G. CMP is also grateful for support from the Departments of Physics at the University of California, Santa Cruz, and the Georgia Institute of Technology, as well as from the DEEP2 survey. JML acknowledges support from the NOAO Leo Goldberg Fellowship and NASA grant HST-AR-9998. We also acknowledge support from the NSF grants AST 05-07483 and AST 08-08133 and the NASA {\it Chandra\/} grants GO5-6141A and GO8-9129A.
We thank the referee and David Koo for very helpful comments.
 This study makes use of data from AEGIS, a multi-wavelength sky survey conducted with the {\it Chandra}, GALEX, {\it Hubble}, Keck, CFHT, MMT, Subaru, Palomar, {\it Spitzer}, VLA and other telescopes and supported in part by the NSF, NASA and the STFC. In addition, this research has made use of the NASA/IPAC Extragalactic Database (NED) which is operated by the Jet Propulsion Laboratory, California Institute of Technology, under contract with the National Aeronautics and Space Administration.

\bsp

\label{lastpage}


\begin{thebibliography}{99} %% list all authors if 8 or fewer
\bibitem[\protect\citeauthoryear{}{}]{b} Aird J.\ et al., 2010, MNRAS, 401, 2531
\bibitem[\protect\citeauthoryear{}{}]{b} Ammons S.\ M., 2009, PhD thesis, Univ.\ of California, Santa Cruz
\bibitem[\protect\citeauthoryear{}{}]{b} Baldry I.\ K., Glazebrook K., Brinkmann J., Ivezi\'{c} $\breve{\rm Z}$., Lupton R.\ H., Nichol R.\ C., Szalay A.\ S., 2004, ApJ, 600, 681
\bibitem[\protect\citeauthoryear{}{}]{b} Baldwin A., Phillips M.\ M., Terlevich R., 1981, PASP, 93, 5
\bibitem[\protect\citeauthoryear{}{}]{b} Barger A.\ J., Cowie L.\ L., Mushotzky R.\ F., Yang Y., Wang W.-H., Steffen A.\ T., Capak P., 2005, AJ, 129, 578
\bibitem[\protect\citeauthoryear{}{}]{b} Barger A.\ J., Cowie L.\ L., Wang W.-H., 2007, ApJ, 654, 764
\bibitem[\protect\citeauthoryear{}{}]{b} Barmby P.\ et al., 2006, ApJ, 642, 126
\bibitem[\protect\citeauthoryear{}{}]{b} Bell E.\ F., de Jong R.\ S., 2001, ApJ, 550, 212
\bibitem[\protect\citeauthoryear{}{}]{b} Bell E.\ F.\ et al., 2004, ApJ, 600, L11
\bibitem[\protect\citeauthoryear{}{}]{b} Bell E.\ F.\ et al., 2005, ApJ, 625, 23
\bibitem[\protect\citeauthoryear{}{}]{b} Bower R.\ G., Benson A.\ J., Malbon R., Helly J.\ C., Frenk C.\ S., Baugh C.\ M., Cole S., Lacey C.\ G., 2006, MNRAS, 370, 645
\bibitem[\protect\citeauthoryear{}{}]{b} Bruzual G., Charlot S., 2003, MNRAS, 344, 1000
\bibitem[\protect\citeauthoryear{}{}]{b} Bundy K.\ et al., 2006, ApJ, 651, 120
\bibitem[\protect\citeauthoryear{}{}]{b} Bundy K.\ et al., 2008, ApJ, 681, 931
\bibitem[\protect\citeauthoryear{}{}]{b} Cattaneo A., Dekel A., Devriendt J., Guiderdoni B., Blaizot J., 2006, MNRAS, 360, 1651
\bibitem[\protect\citeauthoryear{}{}]{b} Ciliegi P., Zamorani G., Hasinger G., Lehmann I., Szokoly G., Wilson G., 2003, A\&A, 398, 901
\bibitem[\protect\citeauthoryear{}{}]{b} Ciotti L., Ostriker J.\ P., 2007, ApJ, 665, 1038
\bibitem[\protect\citeauthoryear{}{}]{b} Coil A.\ L.\ et al., 2009, ApJ, 701, 1484
\bibitem[\protect\citeauthoryear{}{}]{b} Condon J.\ J., 1992, ARA\&A, 30, 575
\bibitem[\protect\citeauthoryear{}{}]{b} Conselice C.\ J.\ et al., 2007, ApJ, 660, L55
\bibitem[\protect\citeauthoryear{}{}]{b} Croton D.\ J.\ et al., 2006, MNRAS, 365, 11
\bibitem[\protect\citeauthoryear{}{}]{b} Davis M.\ et al., 2003, in Guhathakurta P., ed., Proc.\ SPIE Vol.\ 4834, Discoveries and Research Prospects from 6- to 10-Meter-Class Telescopes II. Soc.\ Phot-Opt.\ Instrum.\ Eng., Bellingham, WA, p.\ 161
\bibitem[\protect\citeauthoryear{}{}]{b} Davis M.\ et al., 2007, ApJ, 660, L1
\bibitem[\protect\citeauthoryear{}{}]{b} Dekel A., Birnboim Y., 2006, MNRAS, 368, 2
\bibitem[\protect\citeauthoryear{}{}]{b} Donley J.\ L., Rieke G.\ H., Rigby J.\ R., P\'{e}rez-Gonz\'{a}lez P.\ G., 2005, ApJ, 634, 169
\bibitem[\protect\citeauthoryear{}{}]{b} Drake C.\ L., McGregor P.\ J., Dopita M.\ A., van Breugel W.\ J.\ M., 2003, AJ, 126, 2237
\bibitem[\protect\citeauthoryear{}{}]{b} Faber S.\ M.\ et al. 2007, ApJ, 665, 265
\bibitem[\protect\citeauthoryear{}{}]{b} Fang T.\ et al., 2007, ApJ, 660, L27
\bibitem[\protect\citeauthoryear{}{}]{b} Fasano G., Franceschini A., 1987, MNRAS, 225, 155
\bibitem[\protect\citeauthoryear{}{}]{b} Gehrels N., 1986, ApJ, 303, 336
\bibitem[\protect\citeauthoryear{}{}]{b} Georgakakis A.\ E., Georgantopoulos I., Akylas A., 2006, MNRAS, 366, 171
\bibitem[\protect\citeauthoryear{}{}]{b} Georgakakis A.\ et al., 2007, ApJ, 660, L15
\bibitem[\protect\citeauthoryear{}{}]{b} Georgakakis A., Gerke B.\ F., Nandra K., Laird E.\ S., Coil A.\ L., Cooper M.\ C., Newman J.\ A., 2008a, MNRAS, 391, 183
\bibitem[\protect\citeauthoryear{}{}]{b} Georgakakis A.\ et al., 2008b, MNRAS, 385, 2049
\bibitem[\protect\citeauthoryear{}{}]{b} Georgakakis A.\ et al., 2009, MNRAS, 397, 623

\bibitem[\protect\citeauthoryear{}{}]{b} Gerke B.\ F.\ et al., 2007, ApJ, 660, L23
\bibitem[\protect\citeauthoryear{}{}]{b} Granato G.\ L., De Zotti G., Silva L., Bressan A., Danese L., 2004, ApJ, 600, 580
\bibitem[\protect\citeauthoryear{}{}]{b} Grogin N.\ A.\ et al., 2003, ApJ, 595, 685
\bibitem[\protect\citeauthoryear{}{}]{b} Grogin N.\ A.\ et al., 2005, ApJ, 627, L97
\bibitem[\protect\citeauthoryear{}{}]{b} Hatziminaoglou E.\ et al., 2005, AJ, 129, 1198
\bibitem[\protect\citeauthoryear{}{}]{b} Hopkins P.\ F., Hernquist L., Cox T.\ J., Di Matteo T., Martini P., Robertson B., Springel V., 2005a, ApJ, 630, 705
\bibitem[\protect\citeauthoryear{}{}]{b} Hopkins P.\ F., Hernquist L., Martini P., Cox T.\ J., Robertson B., Di Matteo T., Springel V., 2005b, ApJ, 625, L71
\bibitem[\protect\citeauthoryear{}{}]{b} Hopkins P.\ F., Hernquist L., Cox T.\ J., Robertson B., Springel V., 2006, ApJS, 1635, 50
\bibitem[\protect\citeauthoryear{}{}]{b} Hopkins P.\ F., Hernquist L., Cox T.\ J., Kere$\check{\rm s}$ D., 2008a, ApJS, 175, 356
\bibitem[\protect\citeauthoryear{}{}]{b} Hopkins P.\ F., Cox T.\ J., Kere$\check{\rm s}$ D., Hernquist L., 2008b, ApJS, 175, 390
\bibitem[\protect\citeauthoryear{}{}]{b} Huang J.-S.\ et al., 2007, ApJ, 660, L69
\bibitem[\protect\citeauthoryear{}{}]{b} Ivison R.\ J.\ et al., 2007, ApJ, 660, L77
\bibitem[\protect\citeauthoryear{}{}]{b} Jeltema T.\ E.\ et al., 2009, MNRAS, 399, 715
\bibitem[\protect\citeauthoryear{}{}]{b} Kassin S.\ A.\ et al., 2007, ApJ, 660, L35
\bibitem[\protect\citeauthoryear{}{}]{b} Kauffmann G.\ et al., 2003, MNRAS, 346, 1055
\bibitem[\protect\citeauthoryear{}{}]{b} Kauffmann G.\ et al., 2007, ApJS, 173, 357
\bibitem[\protect\citeauthoryear{}{}]{b} Kennicutt R.\ C., 1998, ARA\&A, 36, 189
\bibitem[\protect\citeauthoryear{}{}]{b} Kinney A.\ L., Calzetti D., Bohlin R.\ C., McQuade K., Storchi-Bergmann T., Schmitt H.\ R., 1996, ApJ, 467, 38
\bibitem[\protect\citeauthoryear{}{}]{b} Konidaris N.\ P.\ et al., 2007, ApJ, 660, L7
\bibitem[\protect\citeauthoryear{}{}]{b} Koo D.\ C.\ et al., 2005, ApJS, 157, 175
\bibitem[\protect\citeauthoryear{}{}]{b} Lacy M.\ et al., 2004, ApJS, 154, 166
\bibitem[\protect\citeauthoryear{}{}]{b} Laird E.\ S., Nandra K., Adelberger K.\ L., Steidel C.\ C., Reddy N.\ A., 2005, MNRAS, 359, 47
\bibitem[\protect\citeauthoryear{}{}]{b} Laird E.\ S.\ et al., 2009, ApJS, 180, 102
\bibitem[\protect\citeauthoryear{}{}]{b} Le Floc'h E.\ et al., 2007, ApJ, 660, L65
\bibitem[\protect\citeauthoryear{}{}]{b} Lin L.\ et al., 2007, ApJ, 660, L51
\bibitem[\protect\citeauthoryear{}{}]{b} Martin D.\ C.\ et al., 2005, ApJ, 619, 1
\bibitem[\protect\citeauthoryear{}{}]{b} Martin D.\ C.\ et al., 2007, ApJS, 173, 342
\bibitem[\protect\citeauthoryear{}{}]{b} Moustakas L.\ A.\ et al., 2007, ApJ, 660, L31
\bibitem[\protect\citeauthoryear{}{}]{b} Nandra K., Pounds K.\ A., 1994, MNRAS, 268, 405
\bibitem[\protect\citeauthoryear{}{}]{b} Nandra K.\ et al., 2005, MNRAS, 356, 568
\bibitem[\protect\citeauthoryear{}{}]{b} Nandra K.\ et al., 2007, ApJ, 660, L11
\bibitem[\protect\citeauthoryear{}{}]{b} Noeske K.\ G.\ et al., 2007a, ApJ, 660, L43
\bibitem[\protect\citeauthoryear{}{}]{b} Noeske K.\ G.\ et al., 2007b, ApJ, 660, L47
\bibitem[\protect\citeauthoryear{}{}]{b} Park S.\ Q.\ et al., 2008, ApJ, 678, 744
\bibitem[\protect\citeauthoryear{}{}]{b} Peterson B.\ M., 1997, An Introduction to Active Galactic Nuclei. Cambridge Univ.\ Press, Cambridge
\bibitem[\protect\citeauthoryear{}{}]{b} Pierce C.\ M.\ et al., 2007, ApJ, 660, L19
\bibitem[\protect\citeauthoryear{}{}]{b} Pierce C.\ M.\ et al., 2010, MNRAS, 405, 718
\bibitem[\protect\citeauthoryear{}{}]{b} Rigby J.\ R., Rieke G.\ H., Donley J.\ L., Alonso-Herrero A., P\'{e}rez-Gonz\'{a}lez P.\ G., 2006, ApJ, 645, 115
\bibitem[\protect\citeauthoryear{}{}]{b} Rovilos E., Georgantopoulos I., 2007, A \& A, 475, 115
\bibitem[\protect\citeauthoryear{}{}]{b} Salim S.\ et al., 2005, ApJ, 619, L39
\bibitem[\protect\citeauthoryear{}{}]{b} Salim S.\ et al., 2007, ApJS, 173, 267
\bibitem[\protect\citeauthoryear{}{}]{b} Salim S.\ et al., 2009, ApJ, 700, 161
\bibitem[\protect\citeauthoryear{}{}]{b} S\'{a}nchez S.\ F.\ et al., 2004, ApJ, 614, 586
\bibitem[\protect\citeauthoryear{}{}]{b} Sato T., Martin C.\ L., Noeske K.\ G., Koo D.\ C., Lotz J.\ M., 2009, ApJ, 696, 214
\bibitem[\protect\citeauthoryear{}{}]{b} Scannapieco E., Silk J., Bouwens R., 2005, ApJ, 635, L13
\bibitem[\protect\citeauthoryear{}{}]{b} Schawinski K., Thomas D., Sarzi M., Maraston C., Kaviraj S., Joo S.-J., Yi S.\ K., Silk J., 2007, MNRAS, 382, 1415
\bibitem[\protect\citeauthoryear{}{}]{b} Schiminovich D.\ et al., 2007, ApJS, 173, 315
\bibitem[\protect\citeauthoryear{}{}]{b} Silk J., Rees M.\ J., 1998, A\&A, 331, L1
\bibitem[\protect\citeauthoryear{}{}]{b} Silverman J.\ D.\ et al., 2005, ApJ, 618, 123
\bibitem[\protect\citeauthoryear{}{}]{b} Silverman J.\ D.\ et al., 2009, ApJ, 696, 396
\bibitem[\protect\citeauthoryear{}{}]{b} Stern D.\ et al., 2005, ApJ, 631, 163
\bibitem[\protect\citeauthoryear{}{}]{b} Symeonidis M.\ et al., 2007, ApJ, 660, L73
\bibitem[\protect\citeauthoryear{}{}]{b} Teng S., Wilson A.\ S., Veilleux S., Young A.\ J., Sanders D.\ B., Nagar N.\ M., 2005, ApJ, 633, 664
\bibitem[\protect\citeauthoryear{}{}]{b} Weiner B.\ J.\ et al., 2007, ApJ, 660, L39  % weiner, papovich, bundy...
\bibitem[\protect\citeauthoryear{}{}]{b} Weiner B.\ J.\ et al., 2009, ApJ, 692, 187  % weiner, coil, prochaska...
\bibitem[\protect\citeauthoryear{}{}]{b} Willmer C.\ N.\ A.\ et al., 2006, ApJ, 647, 853
\bibitem[\protect\citeauthoryear{}{}]{b} Wilson G.\ et al., 2007, ApJ, 660, L59
\bibitem[\protect\citeauthoryear{}{}]{b} Wyder T.\ K.\ et al., 2007, ApJS, 173, 293
\bibitem[\protect\citeauthoryear{}{}]{b} Yan R., Newman J.\ A., Faber S.\ M., Konidaris N., Koo D., Davis M., 2006, ApJ, 648, 281
\end{thebibliography}
\end{document}